\documentclass[preprint,aps,floatfix,nofootinbib,a4paper, superscriptaddress]{revtex4-1}

\usepackage{amsmath, amssymb, amsfonts, amsthm, latexsym, epsfig, mathrsfs, xcolor, bbm, slashed, float}

\usepackage[all]{xy}

\usepackage[inline]{enumitem}

\usepackage{setspace}
\usepackage[marginal, multiple]{footmisc}

\usepackage[utf8]{inputenc}
\usepackage[T1]{fontenc}
\usepackage{lmodern}

\usepackage{microtype}

\usepackage[colorlinks, allcolors=blue!70!black, linktocpage]{hyperref}

\numberwithin{equation}{section}

\usepackage{cleveref}

\let\OLDthebibliography\thebibliography
\renewcommand\thebibliography[1]{%
	\setstretch{1.079} 
	\OLDthebibliography{#1}%
	\small %
	\setlength{\itemsep}{0.2\baselineskip} 
}

\let\OLDfootnote\footnote
\renewcommand\footnote[1]{%
    \count\footins = 1500%
	\setlength{\footnotesep}{0.75\baselineskip}%
	{\footnotesize \OLDfootnote{#1}}%
}

\setlist[enumerate]{noitemsep, label=(\arabic*), ref=(\arabic*)}

\renewcommand\thesection{\arabic{section}}
\renewcommand\thesubsection{\arabic{subsection}}

\makeatletter
\def\p@subsection{\thesection.}
\def\p@subsubsection{\thesection.\thesubsection.}
\makeatother

\theoremstyle{plain}
\newtheorem{thm}{Theorem}
\newtheorem{lemma}{Lemma}[section]

\theoremstyle{definition}

\theoremstyle{remark}
\newtheorem{remark}{Remark}[section]

\crefname{equation}{Eq.}{Eqs.}
\creflabelformat{equation}{#2#1#3}

\crefname{section}{Sec.}{Secs.}
\crefname{appendix}{Appendix}{Appendices}
\crefname{figure}{Fig.}{Figs.}

\crefname{definition}{Def.}{Defs.}
\crefname{prop}{Prop.}{Props.}
\crefname{lemma}{Lemma}{Lemmas}
\crefname{corollary}{Cor.}{Cors.}
\crefname{thm}{Theorem}{Theorems}
\crefname{remark}{Remark}{Remarks}

\crefname{ass}{Assumptions}{Assumptions}
\crefname{property}{Properties}{Properties}

\newcommand{\be}{\begin{equation}}
\newcommand{\ee}{\end{equation}}

\newcommand{\lb}{\left}
\newcommand{\rb}{\right}

\let\oldint\int
\renewcommand{\int}{\oldint\limits}

\newcommand{\lra}{\leftrightarrow}

\newcommand{\mc}{\mathcal}

\newcommand{\ms}{\mathscr}
\newcommand{\mf}{\mathfrak}
\newcommand{\bb}{\mathbb}

\newcommand{\eqsp}{\, ,\quad} 

\newcommand{\pb}[1]{\underleftarrow{#1}}

\newcommand{\hateq}{\mathrel{\mathop {\widehat=} }} 

\newcommand{\Lie}{\pounds} 
\newcommand{\defn}{\mathrel{\mathop:}=} 

\newcommand{\df}[1]{\boldsymbol{#1}} 

\renewcommand{\Re}{{\rm Re}}

\renewcommand{\bar}{\overline}

\newcommand{\scri}{\ms I}

\begin{document}

\setstretch{1.2}


\title{Angular momentum at null infinity in Einstein-Maxwell theory}

\author{B\'eatrice Bonga}
\email{bbonga@perimeterinstitute.ca}
\affiliation{Perimeter Institute for Theoretical Physics, Waterloo, Ontario, N2L 2Y5, Canada}

\author{Alexander M. Grant}
\email{amg425@cornell.edu}
\affiliation{Department of Physics, Cornell University, Ithaca, New York 14853, USA}

\author{Kartik Prabhu}
\email{kartikprabhu@cornell.edu}
\affiliation{Cornell Laboratory for Accelerator-based Sciences and Education (CLASSE),\\ Cornell University, Ithaca, New York 14853, USA}

\begin{abstract}
On Minkowski spacetime, the angular momentum flux through null infinity of Maxwell fields, computed using the stress-energy tensor, depends not only on the radiative degrees of freedom, but also on the Coulombic parts. However, the angular momentum also can be computed using other conserved currents associated with a Killing field, such as the Noether current and the canonical current. The flux computed using these latter two currents is purely radiative. \emph{A priori}, it is not clear which of these is to be considered the ``true'' flux of angular momentum for Maxwell fields. This situation carries over to Maxwell fields on non-dynamical, asymptotically flat spacetimes for fluxes associated with the Lorentz symmetries in the asymptotic Bondi-Metzner-Sachs (BMS) algebra.

We investigate this question of angular momentum flux in the full Einstein-Maxwell theory. Using the prescription of Wald and Zoupas, we compute the charges associated with any BMS symmetry on cross-sections of null infinity. The change of these charges along null infinity then provides a flux. For Lorentz symmetries, the Maxwell fields contribute an additional term, compared to the Wald-Zoupas charge in vacuum general relativity, to the charge on a cross-section. With this additional term, the flux associated with Lorentz symmetries, e.g., the angular momentum flux, is purely determined by the radiative degrees of freedom of the gravitational and Maxwell fields. In fact, the contribution to this flux by the Maxwell fields is given by the radiative Noether current flux and not by the stress-energy flux.
\end{abstract}

\maketitle
\tableofcontents

\section{Introduction}
\label{sec:intro}

There is a surprising fact in Maxwell electromagnetism on Minkowski spacetime. While one typically thinks of fluxes of energy, linear momentum and angular momentum radiated away to null infinity as depending only on the radiative degrees of freedom, this is not always true. While the flux of energy and linear momentum is completely determined by the radiative fields, the flux of angular momentum, when calculated using the stress-energy tensor, also depends on the Coulombic degrees of freedom \cite{ab1,ab2}. These Coulombic degrees of freedom appear through an interaction term with the radiative degrees of freedom and are relevant only if the total charge of the system is nonzero.
This occurs in realistic scenarios: for instance, \textit{all} of the angular momentum radiated by a charged spinning sphere with variable angular velocity is due to the interaction term between radiative and Coulombic degrees of freedom \cite{bpy}.

However, there are other conserved currents for Maxwell fields that are also naturally associated with Killing symmetries in Minkowski spacetime:
\begin{enumerate*}
    \item Using the Lagrangian, one can define a Noether current for Maxwell fields which is the natural conserved current associated with Killing symmetries through Noether's theorem;
    \item Similarly, using the covariant phase space formalism, one can also define a canonical current associated with Killing symmetries.
\end{enumerate*}
Just like the current defined by the stress-energy tensor, each of these currents is conserved, and can be used to define the flux of energy and linear momentum (associated with a time or space translation Killing field) and angular momentum (associated with a rotational Killing field). The fluxes through finite regions of null infinity defined by these conserved currents differ by ``boundary terms'' on the cross-sections bounding this region. When one instead considers the flux through all of null infinity, the difference between these currents depends on the Coulombic part of the Maxwell fields evaluated at spacelike and timelike infinity, which is non-vanishing in general. In particular, in the context of the electromagnetic memory~\cite{Bieri2013}, this difference is nonzero. Thus, \emph{a priori}, it is not obvious which (if any) of these currents defines the ``correct'' notion of energy and angular momentum flux at null infinity for Maxwell fields on Minkowski spacetime.\\

In this paper, we first show that the above considerations generalize to the asymptotic symmetries in Maxwell theory on any non-dynamical, asymptotically flat background spacetime. In particular, one can define the fluxes through null infinity using any of the aforementioned currents associated with the generators of the Bondi-Metzner-Sachs (BMS) algebra. We find that the Noether and canonical currents define fluxes associated with \emph{all} BMS symmetries, and these fluxes are completely determined by the radiative degrees of freedom of the Maxwell fields. However, the flux associated with asymptotic Lorentz symmetries that is defined by the stress-energy current depends also on the Coulombic part via a ``boundary term'' exactly as in Minkowski spacetime. Furthermore, none of these fluxes can be written as the change of a charge computed purely on cross-sections of null infinity. Thus, working purely on null infinity, none of these fluxes can be interpreted as the change in ``energy'' or ``angular momentum'' on cross-sections of null infinity.

To investigate this issue in more detail, we then consider the full Einstein-Maxwell theory, with the background metric now also considered a dynamical field. Unlike Maxwell theory on a non-dynamical background, Einstein-Maxwell theory is diffeomorphism covariant. Thus, we can apply the general prescription of Wald and Zoupas \cite{WZ} to define charges \(\mc Q\) (on any cross-section of null infinity) and their fluxes \(\mc F\) (which are the change in charges \(\mc Q\) through any region of null infinity) associated with the BMS symmetries at null infinity.

We show that if one takes the Wald-Zoupas charges for the BMS symmetries to be defined by the same expression as in vacuum general relativity (say \(\mc Q_{\rm GR}\), \cref{eq:Q-GR-defn}), then the additional contribution to their fluxes due to Maxwell fields is indeed given by the stress-energy current. Consequently, the flux of charges associated with asymptotic Lorentz symmetries, such as angular momentum, is not purely radiative but depends also on the Coulombic parts of the Maxwell fields. However, applying the Wald-Zoupas prescription to the full Einstein-Maxwell theory also gives an additional contribution to the charges themselves due to the Maxwell fields (say \(\mc Q_{\rm EM}\), \cref{eq:Q-EM-defn}). The full Wald-Zoupas charge for Einstein-Maxwell theory is then given by \(\mc Q = \mc Q_{\rm GR} + \mc Q_{\rm EM}\). We show that the flux \(\mc F\) of this full Wald-Zoupas charge across any region of null infinity is completely determined by the radiative degrees of freedom of both the gravitational and Maxwell fields at null infinity. The contribution of the Maxwell fields to this Wald-Zoupas flux is, in fact, given by the Noether current and \emph{not} the stress-energy current. In addition, the Wald-Zoupas flux \(\mc F\) through \emph{all} of null infinity defines a Hamiltonian generator associated with the BMS symmetries on the radiative phase space of Einstein-Maxwell theory at null infinity.

We further show that the additional contribution \(\mc Q_{\rm EM}\) vanishes for supertranslations and does not contribute to the supermomentum charges associated with supertranslation symmetries. In particular, the supermomentum charge is given by the usual formula \(\mc Q_{\rm GR}\) as in vacuum GR, and the supermomentum flux gets an additional (purely radiative) contribution from the Maxwell fields which is equal to the flux determined by the stress-energy or Noether current (as they are equal for supertranslations). If one considers the Kerr-Newman solution, the additional contribution \(\mc Q_{\rm EM}\) vanishes for Lorentz symmetries as well. However, for non-stationary solutions of Einstein-Maxwell theory, \(\mc Q_{\rm EM}\) is generically non-vanishing for Lorentz symmetries. Thus, in general, the contribution due to Maxwell fields to the Wald-Zoupas flux of Lorentz charges, e.g. angular momentum, is not given by the flux of stress-energy but instead by the Noether current flux.\\

The rest of the paper is organized as follows. In \cref{sec:EM}, we review the natural currents of Maxwell theory associated with vector fields in a non-dynamical spacetime which are conserved for Killing vector fields. In \cref{sec:EM-currents-scri}, we consider the limits of these currents to null infinity for BMS vector fields, which need not be exact Killing vector fields, and define the corresponding fluxes associated with the BMS symmetries. In \cref{sec:GR-EM}, we consider Einstein-Maxwell theory, analyze its symplectic current, and review the asymptotic conditions at null infinity. In \cref{sec:WZ}, we consider the Wald-Zoupas prescription to define charges and fluxes associated with the BMS algebra in Einstein-Maxwell theory. We review the essential ingredients of the Wald-Zoupas prescription in \cref{sec:WZ-summ} and compute the charges and fluxes for Einstein-Maxwell theory at null infinity in \cref{sec:WZ-GREM}. We end with \cref{sec:disc} by discussing our main results and their implications. 

Several proofs and explicit computations are relegated to appendices. In \cref{sec:asymp-symm}, we derive useful properties of the asymptotic symmetries of Einstein-Maxwell theory. Some properties of stationary solutions in Einstein-Maxwell theory at null infinity are presented in \cref{sec:stationary}. In \cref{sec:examples}, we collect the computations of the Maxwell contribution to the Wald-Zoupas charge in Kerr-Newman spacetime and for a charged spinning sphere in Minkowski spacetime.

\subsection{Notation and conventions}
\label{sec:notation-and-conventions}

Our notations and conventions are as follows: lowercase Latin indices from the beginning of the alphabet ($a$, $b$, etc.) refer to abstract indices.
Differential forms, when appearing without indices, are in bold.
We follow the conventions of Wald~\cite{Wald-book} for the metric $g_{ab}$, Riemann tensor $R_{abc}{}^d$, and differential forms.
Contraction of vectors into the first index of a differential form is denoted by ``$\cdot$'', e.g. \(X \cdot \df\theta \equiv X^c \theta_{cab}\) for a vector field \(X^a\) and a \(3\)-form \(\df\theta \equiv \theta_{abc}\).

We use the usual conformal completion definition of null infinity $\scri$ with conformal factor $\Omega$ (for a review, see~\cite{Geroch-asymp}). For definiteness we will consider future null infinity --- depending on the conventions some of our formulae will acquire an additional sign when using past null infinity instead. Fields in the physical spacetime are denoted with hats while the corresponding unphysical quantities are unhatted; e.g., \(\hat g_{ab}\) is the physical spacetime metric while \(g_{ab}\) is the metric in the unphysical (conformally-completed) spacetime. The conversion between the metrics and volume elements in the physical and unphysical spacetimes is given by
\begin{equation}\label{eq:g_smooth}
  \hat{g}_{ab} = \Omega^{-2} g_{ab} \eqsp \hat{g}^{ab} = \Omega^2 g^{ab} \eqsp \hat{\varepsilon}_{abcd} = \Omega^{-4} \varepsilon_{abcd}. 
\end{equation}
Let \(n_a \defn \nabla_a \Omega\). It can be shown that the conformal factor \(\Omega\) can always be chosen so that the \emph{Bondi condition}
\be\label{eq:Bondi-cond}
    \nabla_a n_b \hateq 0
\ee
is satisfied, where ``$\hateq$'' denotes equality on $\scri$. Furthermore, with this choice we also have
\be\label{eq:nn-cond}
    n_a n^a = O(\Omega^2).
\ee
We will work with this choice of conformal factor throughout. Let \(q_{ab}\) denote the pullback of the unphysical metric \(g_{ab}\) to \(\scri\). From \cref{eq:Bondi-cond,eq:nn-cond}, it follows that \(q_{ab}n^b \hateq 0\) and \(\Lie_n q_{ab} \hateq 0\). Thus, \(q_{ab}\) defines a degenerate metric on \(\scri\) and a Riemannian metric on the space of null generators (diffeomorphic to \(\bb S^2\)) of \(\scri\).

For our computations, it will be convenient to define some additional structure on \(\scri\) as follows. Let \(u\) be a function on \(\scri\) such that \(n^a\nabla_a u \hateq 1\); i.e., \(u\) is a coordinate along the null generators of \(\scri\) with \(n^a \partial_a \hateq \partial_u\). Consider the foliation of \(\scri\) by a family of cross-sections given by \(u = \text{constant}\). The pullback of \(q_{ab}\) to any such cross-section \(S\) defines a Riemannian metric on \(S\). For such a choice of foliation, there is a unique \emph{auxiliary normal} vector field \(l^a\) at \(\scri\) such that
\be\label{eq:l-props}
    l^a l_a \hateq 0 \eqsp l^a n_a \hateq -1 \eqsp q_{ab}l^b \hateq 0.
\ee
Note that this choice of auxiliary normal is parallel-transported along \(n^a\), i.e. \(n^b \nabla_b l^a \hateq 0\).\footnote{All of our results can be obtained without choosing a foliation of \(\scri\) and the corresponding auxiliary normal \(l^a\), but some intermediate computations become more cumbersome; see \cite{Geroch-asymp,Ash-Str}.}

In terms of this auxiliary normal, we also have
\be\label{eq:null-fields}
    q_{ab} \hateq g_{ab} + 2 n_{(a} l_{b)} \eqsp q^{ab} \hateq g^{ab} + 2 n^{(a} l^{b)}.
\ee
where \(q^{ab}\) is the ``inverse metric'' on the chosen foliation relative to \(l^a\). For any \(v_a\) satisfying \(n^a v_a \hateq l^a v_a \hateq 0\) on \(\scri\), we define the derivative \(\ms D_a\) on the cross-sections by
\be
    \ms D_a v_b \defn q_a{}^c q_b{}^d \nabla_c v_d.
\ee
It is easily verified that \(\ms D_a q_{bc} \hateq 0\); i.e., \(\ms D_a\) is the metric-compatible covariant derivative on cross-sections of \(\scri\).

Let $\df \varepsilon_3 \equiv \varepsilon_{abc}$ be the volume element on \(\scri\) and $\df \varepsilon_2 \equiv \varepsilon_{ab}$ the area element on the cross-sections of \(\scri\) in our choice of foliation which we define by
\be\label{eq:volumes}
    \varepsilon_{abc} \defn l^d \varepsilon_{dabc} \eqsp \varepsilon_{ab} \defn - n^c \varepsilon_{cab} \, .
\ee
These are the orientations of $\df \varepsilon_3$ and $\df \varepsilon_2$ that are used by~\cite{WZ}. In our choice of foliation, we also have $\df \varepsilon_3 = - d u \wedge \df\varepsilon_2$.

We also use the following terminology for the charges and fluxes associated with the symmetry algebra at null infinity. Quantities associated with asymptotic symmetries evaluated as integrals over cross-sections \(S \cong \bb S^2\) of null infinity will be called ``charges'', while those evaluated as an integral over a portion \(\Delta\scri\) of null infinity bounded by two cross-sections will be called ``fluxes''. In general, fluxes need not be the difference of any charges on the two bounding cross-sections, but the Wald-Zoupas fluxes (defined in \cref{sec:WZ}) are the change of the Wald-Zoupas charges. When certain conditions are satisfied the fluxes given by the Wald-Zoupas prescription can also be considered as Hamiltonian generators on the phase space at null infinity (see the discussion below \cref{eq:hamiltonian}).

\section{Maxwell fields on a non-dynamical background spacetime}
\label{sec:EM}

In this section, we discuss in detail three currents that occur in the theory of Maxwell fields associated with vector fields on a fixed, non-dynamical background spacetime: the canonical, stress-energy and Noether currents. We show that, when the vector field is a Killing field of the background metric, each of these currents is conserved and they differ by ``boundary'' terms. Next, we carefully analyze the fluxes through $\scri$ defined by each of these currents when the vector fields are asymptotic symmetries in the BMS algebra. This serves as a primer for the remaining part of the paper where we analyze Einstein-Maxwell theory at \(\scri\) and define charges and fluxes for its asymptotic symmetries.\\

The dynamical field of Maxwell electrodynamics is given by a vector potential. It is most natural to treat the vector potential as a connection on a \(U(1)\)-principal bundle over spacetime, and perform the analysis directly on the principal bundle \cite{KP-bundle}. Since this would need considerable additional formalism, we will instead treat the vector potential as a \(1\)-form \(\hat A_a\) on spacetime which is obtained from the connection by making an (arbitrary) choice of gauge. The Maxwell field strength \(2\)-form \(\hat F_{ab}\) is then
\be
    \hat F_{ab} \defn 2\hat \nabla_{[a} \hat A_{b]}.
\ee
To define our currents we will consider the transformations of the vector potential under both Maxwell gauge transformations parametrized by a function \(\hat \lambda\) and diffeomorphisms generated by a vector field \(\hat X^a\), which we collectively denote by \(\hat\xi = (\hat X^a, \hat \lambda)\). The infinitesimal change in the vector potential under these transformations is given by
\begin{equation}\label{eq:A-transform}
  \delta_{\hat \xi} \hat{A}_a = \Lie_{\hat X} \hat{A}_a + \hat{\nabla}_a \hat{\lambda} = \hat{X}^b \hat{F}_{ba} + \hat{\nabla}_a \left(\hat{X}^b \hat{A}_b + \hat{\lambda}\right).
\end{equation}

Note that the vector field \(\hat X^a\) and the function \(\hat\lambda\) are independent of any choice of gauge for the Maxwell vector potential, since they are simply vector fields and functions on the spacetime. However, for a \emph{fixed} transformation parametrized by \(\hat\xi\), its representation in terms of a vector field \(\hat X^a\) and a Maxwell gauge transformation \(\hat\lambda\) depends on the choice of gauge for the vector potential \(\hat A_a\). Let \(\hat A'_a = \hat  A_a + \hat \nabla_a \hat \Lambda\) be another vector potential related to \(\hat A_a\) by a gauge transformation \(\hat \Lambda\). For a fixed \(\hat \xi = (\hat X^a, \hat \lambda)\) let the new representatives under the gauge transformation by \(\hat \Lambda\) be \(\hat \xi = (\hat X'{}^a, \hat \lambda')\). Since \(\hat \xi\) is fixed, its action on the vector potentials must be independent of the choice of gauge; that is, \(\delta_{\hat \xi} \hat A'_a = \delta_{\hat \xi} \hat  A_a\). Evaluating this, we have
\be\begin{aligned}
    \Lie_{\hat X'} \hat A_a + \hat \nabla_a \hat \lambda' + \hat \nabla_a \Lie_{\hat X'} \hat \Lambda &= \Lie_{\hat X} \hat  A_a + \hat \nabla_a \hat \lambda \; .
\end{aligned}\ee
This implies that, under a change of Maxwell gauge by \(\hat \Lambda\), the representation of a fixed transformation \(\hat \xi = (\hat X^a, \hat \lambda) = (\hat X'{}^a, \hat \lambda')\) changes as
\be\label{eq:symm-gauge}
    \hat X' {}^a = \hat X^a \eqsp \hat \lambda' = \hat \lambda - \Lie_{\hat X} \hat \Lambda \; .
\ee
Consequently, the notion of a pure Maxwell gauge transformation \(\hat \xi = (\hat X^a = 0, \hat \lambda)\) is well-defined independently of the choice of gauge \(\hat \Lambda\), but a ``pure diffeomorphism'' \(\hat \xi = (\hat X^a, \hat \lambda = 0)\) is not. This is analogous to the structure of the BMS algebra noted in \cref{sec:asymp-symm}.  Note also that
\be\label{eq:inv-phys}
    \hat \lambda' + \hat X'{}^a \hat A'_a = \hat \lambda + \hat X^a \hat A_a
\ee
is invariant under changes of Maxwell gauge.\footnote{On a principal bundle, \(\hat\xi=(\hat X^a, \hat\lambda)\) is a vector field on the bundle and \cref{eq:A-transform} is the Lie derivative of the connection with respect to \(\hat \xi\). The Lie algebra of such vector fields also has the structure of a semidirect sum of diffeomorphisms with the Lie ideal of Maxwell gauge transformations \cite{KP-bundle}. The invariant in \cref{eq:inv-phys} is then the vertical part of \(\hat \xi\) on the bundle.}\\

The Lagrangian \(4\)-form of Maxwell electrodynamics is given by
\begin{equation}\label{eq:L-EM}
  \df L_{\rm EM} \defn \hat{\df \varepsilon}_4 \left(-\frac{1}{16\pi} \hat{F}^2 \right),
\end{equation}
where \(\hat F^2 \defn \hat{g}^{ac} \hat{g}^{bd} \hat F_{ab} \hat F_{cd}  \) and the metric is considered to be a non-dynamical field. One can also consider the Maxwell field coupled to a charged source current of compact support. On Minkowski spacetime, such source currents are necessary to have a non-vanishing Coulombic part of the Maxwell field. Of course, there are asymptotically flat spacetimes which are solutions of the source-free Maxwell equations and have a non-vanishing Coulombic part without introducing external sources, e.g., the Kerr-Newman spacetimes. Since we are mostly concerned with the behaviour at null infinity, a source current of compact support does not change our main analysis. However, we assume the presence of such sources to enrich our class of solutions so that also on Minkowski spacetime there exist Maxwell field configurations with a nonzero total charge.

Varying the Lagrangian with respect to the dynamical field \(\hat A_a\) gives
\begin{equation} \label{eq:M_variation-vector-potential}
\delta \df L_{\rm EM} = \hat{\df \varepsilon}_4 \left[\frac{1}{4\pi} \left(\hat{\nabla}_b \hat{F}^{ba}\right) \delta \hat{A}_a - \frac{1}{4\pi} \hat{\nabla}_b \left(\hat{F}^{ba} \delta \hat{A}_a\right)\right],
\end{equation}
which yields the Maxwell equations
\begin{equation}
  \hat{\nabla}_b \hat{F}^{ba} = 0,
\end{equation}
as well as a ``boundary term'' corresponding to the symplectic potential \(3\)-form
\be\label{eq:theta-EM}
    \df\theta_{\rm EM}(\delta \hat A) \equiv -\frac{1}{4\pi} \hat\varepsilon_{dabc} \hat{F}^{de} \delta \hat{A}_e.
\ee
The symplectic current \(3\)-form is then defined as
\begin{equation} \label{eq:M_symp_current}
  \df\omega_{\rm EM} \defn \delta_1 \df\theta_{\rm EM}(\delta_2 \hat A) - \delta_2 \df\theta_{\rm EM}(\delta_1 \hat A) \equiv
    -\frac{1}{4\pi} \hat\varepsilon_{dabc} \left[\delta_1 \hat{F}^{de} \delta_2 \hat{A}_e - (1 \lra 2)\right].
\end{equation}
From this symplectic current, we construct the \emph{canonical current} for a transformation of the vector potential (\cref{eq:A-transform}) generated by \(\hat\xi = (\hat X^a, \hat\lambda)\). \emph{A priori}, one may naively expect the canonical current to involve two variations of the vector potential. However, since the Maxwell equations are linear, the situation simplifies: consider a one-parameter family of vector potentials $\hat{A}_a (\epsilon) \defn (1 + \epsilon) \hat{A}_a$. This entire family satisfies the Maxwell equations if $\hat{A}_a$ satisfies the Maxwell equations, and the variation of this family of solutions $\delta\hat{A}_a \defn \frac{d}{d \epsilon} \hat{A}_a (\epsilon) \vert_{\epsilon = 0}$ is equal to the vector potential $\hat{A}_a$. Therefore, for a given symmetry $\hat{\xi} \defn (\hat{X}^a, \hat{\lambda})$, where $\hat{X}^a$ is any vector field and $\hat{\lambda}$ is gauge, we define the \emph{canonical current} as

\begin{equation} \label{eq:M_canonical}\begin{aligned}
\df J_{\rm C}[\hat\xi] &\defn \df\omega_{\rm EM}(\hat A, \delta_{\hat\xi}\hat A) \equiv \hat\varepsilon_{dabc} \hat{\jmath}_{\rm C}^d \\
\text{with}\quad
\hat{\jmath}_{\rm C}^a &= -\frac{1}{4\pi} \left[\hat{F}^{ab} \left(\Lie_{\hat X} \hat{A}_b + \hat{\nabla}_b \hat{\lambda}\right) - \hat{g}^{ac} \hat{g}^{bd} \hat{A}_b \Lie_{\hat X} \hat{F}_{cd} \right] \; .
\end{aligned}\end{equation}\\

To define the stress-energy and Noether currents, we also need to vary the Maxwell Lagrangian with respect to the metric \(\hat g_{ab}\).%
\footnote{
Note that varying the Lagrangian with respect to $\hat{g}_{ab}$ is not in contradiction with our assumption of $\hat{g}_{ab}$ being non-dynamical in this section --- $\hat{g}_{ab}$ does not satisfy any equation of motion obtained by varying the purely Maxwell Lagrangian.
}
In particular, by varying the Lagrangian with respect to the non-dynamical metric $\hat{g}_{ab}$ we find the Maxwell stress-energy tensor \(\hat T^{ab}\):
\begin{equation} \label{eq:M_variation-metric}
  \delta_{\hat{g}} \df L_{\rm EM} = \hat{\df \varepsilon}_4 \tfrac{1}{2} \hat{T}^{ab} \delta \hat{g}_{ab},
\end{equation}
where
\begin{equation} \label{eq:M_T}
  \hat{T}^{ab} \defn \frac{1}{4\pi} \left(\hat{F}^{ac} \hat{F}^b{}_c - \tfrac{1}{4} \hat{g}^{ab} \hat{F}^2\right).
\end{equation}
The associated current, the \emph{stress-energy current} for some vector field \(\hat X^a\), is given by
\begin{equation}\begin{aligned}
    \df J_{\rm T} &\equiv \hat\varepsilon_{dabc} \hat{\jmath}_{\rm T}^d, \\
    \text{with}\quad
  \hat{\jmath}_{\rm T}^a (\hat{X}) &\defn \hat{T}^{ab} \hat{X}_b = \frac{1}{4\pi} \left(\hat{F}^{ac} \hat{F}_{bc} \hat{X}^b - \tfrac{1}{4} \hat{X}^a \hat{F}^2\right).
\end{aligned}\end{equation}
Given that its divergence is
\begin{equation}
  \hat{\nabla}_a \hat{\jmath}_{\rm T}^a (\hat{X}) = \hat{T}^{ab} \hat{\nabla}_{(a} \hat{X}_{b)},
\end{equation}
it is clear that $\hat{\jmath}_{\rm T}^a (\hat{X})$ is conserved when $\hat{X}^a$ is Killing.

We finally turn to the Noether current. To obtain its expression, we consider the variation of the Maxwell Lagrangian under the transformation generated by \(\hat\xi = (\hat X^a, \hat\lambda)\), where the vector potential transforms as in \cref{eq:A-transform} and the variation of the metric under diffeomorphisms is $\delta_{\hat \xi} \hat{g}_{ab} = \Lie_{\hat X} \hat{g}_{ab}$ (see the appendix of \cite{iyer-wald}). This yields\footnote{Note that when the vector field \(\hat X^a\) is non-vanishing it is essential that the non-dynamical metric in the Maxwell Lagrangian is also varied so that $\delta_{\hat \xi} \df L_{\rm EM}$ is a total derivative.}
\begin{equation}
    \delta_{\hat \xi} \df L_{\rm EM} = \Lie_{\hat X} \df L_{\rm EM} = d \df\eta[\hat\xi],
\end{equation}
where the \(3\)-form \(\df\eta[\hat\xi]\) is given by
\begin{equation}\label{eq:eta-defn}
    \df\eta[\hat\xi] = \hat X \cdot \df L_{\rm EM} = -\tfrac{1}{16 \pi} \hat\varepsilon_{dabc} \hat{F}^2 \hat{X}^d.
\end{equation}
The \emph{Noether current} is then defined by (see the appendix of \cite{iyer-wald})
\begin{equation}\label{eq:M_noether}\begin{aligned}
    \df J_{\rm N}[\hat\xi] &\defn \df\theta_{\rm EM}(\delta_{\hat\xi}\hat A_a) - \df\eta[\hat\xi] \equiv \hat\varepsilon_{dabc} \hat{\jmath}_{\rm N}^d, \\
    \text{with}\quad
    \hat{\jmath}_{\rm N}^a &= -\frac{1}{4\pi} \hat{F}^{ab} \left[\Lie_{\hat X} \hat A_b + \hat{\nabla}_b \hat{\lambda}\right] + \frac{1}{16 \pi} \hat{X}^a \hat{F}^2.
\end{aligned}\end{equation}\\

Despite the fact that these three currents are clearly different, in the case where the vector field $\hat{X}^a$ is Killing, all these currents differ only by total derivatives and constant factors. It can be shown quite generally that the Noether and stress-energy currents are related by a total derivative, see the appendix of \cite{iyer-wald}. For Maxwell fields, we find by comparing the Noether and stress-energy current that
\begin{equation} \label{eq:noether_boundary}
  \df J_{\rm N} [\hat{\xi}] = - \df J_{\rm T} [\hat{X}] + d \df Q_{\rm N}[\hat{\xi}],
\end{equation}
where
\begin{equation}\label{eq:Q-N-EM}
  \df Q_{\rm N}[\hat{\xi}] \equiv -\tfrac{1}{8\pi} \hat\varepsilon_{cdab} \hat{F}^{cd} \left(\hat{X}^e \hat{A}_e + \hat{\lambda}\right).
\end{equation}
Comparing the canonical with the Noether current, one instead finds (after a lengthy but straightforward calculation starting with \cref{eq:M_canonical}) that
\begin{equation} \label{eq:canonical_boundary}
  \df J_{\rm C} [\hat{\xi}] = 2 \df J_{\rm N} [\hat{\xi}] + d \df Q_{\rm C}[\hat{\xi}] + \df K_{\rm C},
\end{equation}
where
\begin{align}
  \df Q_{\rm C}[\hat{\xi}] &\defn -\tfrac{1}{8\pi} \hat\varepsilon_{cdab} \left(2 \hat{X}^{c} \hat{F}^{de} \hat{A}_e - \hat{\lambda} \hat{F}^{cd}\right) \\
  \df K_{\rm C} &\defn \tfrac{1}{2\pi} \hat\varepsilon_{dabc} \left(2 \hat{g}^{f[d} \hat{F}^{e]g} - \tfrac{1}{2} \hat{F}^{de} \hat{g}^{fg}\right) \hat{A}_e \hat{\nabla}_{(f} \hat{X}_{g)}.
\end{align}
When $\hat{X}^a$ is a Killing vector field of the background spacetime, the Noether and canonical current differ only by a total derivative of $\df Q_{\rm C} [\hat{\xi}]$ (up to a constant factor of two).\\

For any Killing vector field $\hat{X}^a$, these currents are all related by total derivatives, and the fact that the stress-energy current is conserved in this case directly shows that the other two currents are also conserved. From the discussion under \cref{eq:A-transform}, it follows that both the stress-energy and Noether current are invariant under Maxwell gauge transformations, while the canonical current is invariant only up to boundary terms. Thus we can use any of these currents to define a conserved quantity for Maxwell fields associated with a Killing vector field of the background spacetime.\footnote{Of course, one is free to define other conserved currents by simply adding exact \(2\)-forms (i.e boundary terms) to the three currents we have defined.} For example, if the background spacetime is stationary with a timelike Killing field \(\hat t^a\), then any of the above defined currents with \(\hat X^a = \hat t^a\), integrated over a Cauchy surface, defines a notion of ``energy''. Similarly, for an axisymmetric background with an axial Killing field \(\hat X^a = \hat\phi^a\), each of these currents define an ``angular momentum''. The conserved quantities defined using these currents will then differ by boundary terms on the Cauchy surface, either at a boundary at infinity, or some interior boundary like a black hole horizon.

The most appropriate current to use depends on the problem at hand. The Noether current is the most natural one associated with a symmetry through Noether's theorem (and, as we will show, is also the contribution due to the Maxwell fields to the Wald-Zoupas flux). On the other hand, the stress-energy current is typically used for calculations of energy and angular momentum flux, both in standard textbooks for Maxwell theory in flat spacetimes~\cite{Griffiths,Jackson} and on fixed backgrounds~\cite{Wald-book} (in fact, problem 9.8 of~\cite{Jackson} notes that the angular momentum flux depends on more than just the radiative electromagnetic fields!). Furthermore, for computations of ``self-force'' effects on charged sources due to electromagnetic radiation, the useful quantity to use is the stress-energy current; see, for instance \cite{QW,bpy}.

The canonical currents are associated directly to the Hamiltonian formulation where the symplectic current provides a natural symplectic form on the phase space. These currents also arise in the formulation of the first law of black hole mechanics \cite{iyer-wald,KP-bundle}. By general arguments, the positivity of the canonical energy (relative to a timelike Killing field of the background) is also directly related to the stability of the background black hole to perturbations \cite{HW-stab,PW}. For axisymmetric Maxwell fields on a stationary (but not static) and axisymmetric black hole spacetime in GR, it was shown in \cite{PW-em} that the energy evaluated on a Cauchy surface defined by the canonical current (which, in this case, also equals the one defined by Noether current) is, in fact positive whereas the energy given by the stress-energy tensor can be made negative. Thus, the canonical energy is the more useful quantity in the analysis of stability of black hole spacetimes to electromagnetic perturbations. The canonical energy is also useful to account for the ``second-order'' self-force effects of small test bodies in black hole spacetimes \cite{Sorce-Wald}. Similarly, the symplectic current is useful for deriving conserved currents associated with symmetries of the equations of motion which need not arise from the action of a diffeomorphism or gauge transformation \cite{GF1,GF2}.

\subsection{Maxwell currents and fluxes at $\scri$}
\label{sec:EM-currents-scri}

We now turn to comparing the fluxes through $\scri$ constructed from the various currents in the previous section. Hereafter, we will not require that the vector field \(\hat X^a\) is a Killing field, but instead require it to be an element of the asymptotic BMS symmetry algebra at \(\scri\).  In order to make this comparison, we first list the asymptotic properties of the relevant fields at null infinity.

As usual, we perform this calculation in the unphysical spacetime. The unphysical Maxwell field tensor is given by \(F_{ab} = \hat F_{ab}\), and we assume that \(F_{ab}\) extends smoothly to \(\scri\). For the vector potential, this implies that there exists a gauge in which \(A_a = \hat A_a\) is also smooth at \(\scri\).\footnote{Generically, if we impose some gauge condition on \(\hat A_a\) in the physical spacetime, e.g., Lorenz gauge, then \(A_a = \hat A_a\) is not guaranteed to be smooth at \(\scri\) in the chosen gauge; see, for example, the case of Kerr-Newman spacetime in \cref{sec:examples}.} Moreover, without loss of generality --- that is, for all solutions of the Maxwell equations where \(F_{ab}\) is smooth at \(\scri\) --- we can further restrict the gauge freedom to the \emph{outgoing radiation gauge}
\begin{equation} \label{eq:A_bondi}
  n^a A_a \hateq 0 \, .
\end{equation}
The argument is similar to the one used for imposing the Bondi condition (see for instance, Sec.~11.1 of \cite{Wald-book}): Let \(A_a\) be a vector potential so that \(n^a A_a \not\hateq 0\), and consider another vector potential \(A'_a\) related to it by a Maxwell gauge transformation \(A'_a = A_a + \nabla_a \lambda \). Now choose \(\lambda\) to be a solution of
\be
    \Lie_n \lambda \hateq - n^a A_a \,.
\ee
Since this is an ordinary differential equation along the generators of \(\scri\), solutions to this equation always exist. With this choice of \(\lambda\), we have \(n^a A'_a \hateq 0\). Henceforth, we will assume that this choice has been made for the vector potential.

Now consider a diffeomorphism \(\hat X^a\) and a Maxwell gauge transformation \(\hat \lambda\). We show in \cref{sec:asymp-symm} that to preserve the asymptotic-flatness conditions on the spacetime, \(X^a = \hat X^a\) must be smooth at \(\scri\) and correspond to an element of the BMS Lie algebra. The essential conditions on \(X^a\) at \(\scri\) are collected in \cref{eq:bms-cond,eq:alpha-cond,eq:BMS-decomp}. Similarly, for the transformation of the vector potential (\cref{eq:A-transform}) to preserve our conditions on the Maxwell field we must have that \(\lambda = \hat \lambda\) is smooth at \(\scri\) and satisfies \(\Lie_n \lambda \hateq 0\).

In summary, we have that
\begin{gather}
  \hat A_a = A_a \eqsp \hat{F}_{ab} = F_{ab} \eqsp \hat{X}^a = X^a \eqsp \hat{\lambda} = \lambda. \label{eq:misc_smooth}
\end{gather}
are all smooth at \(\scri\) along with the condition \cref{eq:A_bondi}.

Two important quantities can be derived from the ``electric field'' $F_{ab} n^b$ at $\scri$: the first is $\mc E_a$, defined by
\begin{equation}
  \mc E_a \defn \pb{F_{ab} n^b} = q_a{}^c F_{cb}n^b = - \Lie_n \pb{A_a},
\end{equation}
with the under arrow indicating the pullback to $\scri$.
The radiative degrees of freedom in the electromagnetic field are contained in $\mc E_a$ (or, equivalently, $\pb{A_a}$).
The other piece of $F_{ab} n^b$, which contains non-radiative (Coulombic) information at $\scri$, is given by $\Re[\varphi_1]$, defined by\footnote{The notation ``$\Re[\varphi_1]$'' comes from Newman-Penrose notation~\cite{NP}. Similarly, the quantity \(\mc E_a\) corresponds to the real and imaginary parts of \(\varphi_2\) in Newman-Penrose notation.}
\begin{equation}
  \Re[\varphi_1] \defn \tfrac{1}{2} F_{ab} l^a n^b.
\end{equation}
The Maxwell equations imply that on $\scri$ these two fields are related in the following way:
\be\label{eq:n_maxwell}
  2 \Lie_n \Re[\varphi_1] \hateq q^{ab} \ms D_a \mc E_b.
\ee\\

With these asymptotic conditions we now evaluate the fluxes through null infinity defined by the canonical, Noether, and stress-energy currents for any asymptotic symmetry \(\xi = (X^a, \lambda)\) as described above. Note that in this context the vector field \(\hat X^a = X^a\) need not be a Killing vector field inside the physical spacetime but is required to be a BMS vector field on \(\scri\).

With our convention in \cref{sec:notation-and-conventions} for $\df \varepsilon_3$, the pullback of a \(3\)-form \(\df J\) is \(- \Omega^{-4} n_a \hat \jmath^a ~\df\varepsilon_3\), where $J_{abc} = \hat{\varepsilon}_{abcd} \hat{\jmath}^d$.
The flux of the canonical current is given by
\begin{equation} \label{eq:F_C}
  \begin{split}
    \mc F_{\rm C} [\xi; \Delta\scri] &\defn \int_{\Delta\scri} \df J_{\rm C} [\xi] = - \int_{\Delta\scri} \df\varepsilon_3~ \Omega^{-4} n_a \hat\jmath_{\rm C}^a [\xi] \\
    &= -\frac{1}{4\pi} \int_{\Delta\scri} \df \varepsilon_3 \; q^{ab} \left[\mc E_a (\Lie_X A_b + \mathscr{D}_b \lambda) - A_a \Lie_X \mc E_b - \tfrac{1}{2} \mc E_a A_b \mathscr{D}_c Y^c\right],
  \end{split}
\end{equation}
where $Y^a$ is the ``pure Lorentz part'' of $X^a$ and we have used that $\Lie_X n^a \hateq -\tfrac{1}{2}(\mathscr{D}_b Y^b) n^a$ (see \cref{eq:bms-cond} and the text below \cref{eq:lie-f}).
The flux of the Noether current is given by
\be\label{eq:flux-N}
  \mc F_{\rm N}[\xi; \Delta\scri] \defn \int_{\Delta\scri} \df J_{\rm N} [\xi] = - \int_{\Delta\scri} \df\varepsilon_3~ \Omega^{-4} n_a \hat\jmath_{\rm N}^a [\xi] = -\frac{1}{4\pi}\int_{\Delta\scri} \df\varepsilon_3~ q^{ab} \mc E_a (\Lie_X A_b + \ms D_b \lambda),
\ee
where we have used that $\Lie_n \lambda \hateq 0$ (see \cref{eq:lambda-dot}). The term proportional to $F^2$ in \cref{eq:M_noether} does not contribute to the flux through $\scri$ because $X^a n_a \hateq 0$.
Finally, the flux of the stress-energy current is given by
\be
  \mc F_{\rm T} [\xi; \Delta\scri] \defn \int_{\Delta\scri} \df J_{\rm T} [\xi] = - \int_{\Delta \scri} \df\varepsilon_3~ T_{ab} n^a X^b = - \frac{1}{4\pi}\int_{\Delta\scri} \df\varepsilon_3~ \mc E_a \lb( q^{ab} F_{bc} X^c + 2 \Re[\varphi_1] Y^a\rb) \; .
\ee

From the above expressions it is apparent that all of these fluxes vanish in the absence of electromagnetic radiation, i.e., when \(\mc E_a = 0\). Furthermore, the fluxes defined by the Noether and canonical currents depend only on the radiative modes \(\pb A_a\) at null infinity.
However, the stress-energy current flux also depends on the Coulombic part \(\Re[\varphi_1]\), as emphasized before in \cite{ab1,ab2}.
For supertranslations \(X^a \propto n^a\), this Coulombic term does not contribute to the flux since \(Y^a = 0\).
However, the stress-energy current flux associated with asymptotic Lorentz symmetries, e.g. angular momentum flux, cannot be computed from just the radiative modes.

Note that, since any BMS vector field satisfies \(\Omega^2 \Lie_X \hat g_{ab} \hateq 0\) (see the discussion in \cref{sec:asymp-symm}), the \(3\)-form term \(\df K_{\rm C}\) in \cref{eq:canonical_boundary} vanishes at null infinity.
Thus, from \cref{eq:canonical_boundary} we have on \(\scri\)
\begin{equation}
  \df J_{\rm N} [\xi] \hateq \tfrac{1}{2} \left[\df J_{\rm C}[\xi] - d \df Q_{\rm C} [\xi] \right] \eqsp \df J_{\rm T} [\xi] \hateq -\df J_{\rm N} [\xi] + d \df Q_{\rm N} [\xi].
\end{equation}
That is, all three currents evaluated on \(\scri\) differ by exact \(3\)-forms even when the vector field \(X^a\) is not Killing but an element of the BMS algebra. Therefore, the fluxes of these currents on \(\scri\) can be related to each other purely by boundary terms on the cross-sections \(S_2\) and \(S_1\) bounding the region \(\Delta\scri\) (with $S_2$ in the future of $S_1$).

Let us compare the fluxes on $\scri$ in more detail.
Consider, first, the relation between the flux of the Noether and canonical current.
This satisfies
\begin{equation}
  \mc F_{\rm N}[\xi; \Delta\scri] \defn \int_{\Delta\scri} \df J_{\rm N} (\xi) = \frac{1}{2} \mc F_{\rm C} [\xi; \Delta\scri] + \frac{1}{2} \lb[\int_{S_2} \df Q_{\rm C} [\xi] - \int_{S_1} \df Q_{\rm C} [\xi]\rb], \\
\end{equation}
with the boundary term
\begin{equation} \label{eq:Q_C_integral}
  \int_S \df Q_{\rm C} [\xi] = -\frac{1}{4 \pi} \int_S \df \varepsilon_2 \left(\beta \mc E^a A_a - 2 \lambda \; \Re[\varphi_1]\right) \, ,
\end{equation}
where \(\beta\) is as given in \cref{eq:BMS-decomp}.
This expression is rather strange on first inspection, since both $\mc F_{\rm C}$ and $\mc F_{\rm N}$ contain only radiative information by \cref{eq:F_C,eq:flux-N}, respectively, and yet their difference appears to be a boundary term that contains non-radiative information, in the form of $\lambda \; \Re[\varphi_1]$.
This is somewhat misleading, since using the Maxwell equation \cref{eq:n_maxwell} and $\Lie_n \lambda \hateq 0$, this Coulombic contribution can be rewritten in terms of purely radiative degrees of freedom as
\begin{align}
	\frac{1}{4 \pi} \int_{S_2} \df  \varepsilon_2  \; 2 \lambda \Re [\varphi_1] - \frac{1}{4 \pi} \int_{S_1} \df  \varepsilon_2  \; 2 \lambda \Re [\varphi_1] =
	\frac{1}{4 \pi} \int_{\Delta \scri} \df \varepsilon_3  \; q^{ab} \mc E_a \mathscr{D}_b \lambda \; .
\end{align}

Next, consider the relation between the flux of the stress-energy and Noether current:
\begin{equation}\label{eq:F_N}
  \mc F_{\rm T} [\xi; \Delta\scri] = -\mc F_{\rm N} [\xi; \Delta\scri] - \left[\int_{S_2} \df Q_{\rm N} [\xi] - \int_{S_1} \df Q_{\rm N} [\xi]\right],
\end{equation}
with
\begin{equation} \label{eq:Q_N_value}
  \int_S \df Q_{\rm N} [\xi] = -\frac{1}{2\pi} \int_S \df\varepsilon_2 \Re[\varphi_1] \left(Y^a A_a + \lambda\right).
\end{equation}
Unsurprisingly, as there is non-radiative information in $\mc F_{\rm T}$ but not in $\mc F_{\rm N}$, the boundary term contains non-radiative information.

Finally, let us consider the fluxes through all of $\scri$.
The natural boundary conditions for the electromagnetic field in the limit $u \to \pm \infty$ are
\be
    \mc E_a = O(1/|u|^{1+\epsilon}) \eqsp \pb{A_a} = O(1) \; .
\ee
These conditions ensure that the symplectic form obtained by integrating the symplectic current over \emph{all} of \(\scri\) is finite. Given that \(\beta\) grows at most linearly in \(u\) and \(Y^a\) and $\lambda$ are independent of $u$ (see \cref{sec:asymp-symm}), we find that the fluxes differ by
\begin{align}
  \mc F_{\rm N} [\xi;\scri] &= \tfrac{1}{2} \mc F_{\rm C} [\xi; \scri] + \tfrac{1}{2} \left[\mc Q_{\rm C} (S_\infty) - \mc Q_{\rm C} (S_{-\infty})\right], \\
  \mc F_{\rm T}[\xi; \scri] &= - \mc F_{\rm N}[\xi; \scri] - \left[\mc Q_{\rm N} (S_\infty) - \mc Q_{\rm N} (S_{-\infty})\right],
\end{align}
where $S_\infty$ and $S_{-\infty}$ are the spheres at $u = \pm \infty$, respectively, and
\begin{align}
  \mc Q_{\rm C} (S) &\defn \frac{1}{2\pi} \int_S \df\varepsilon_2~ \lambda \Re[\varphi_1], \\
  \mc Q_{\rm N} (S) &\defn -\frac{1}{2\pi} \int_S \df\varepsilon_2~ \Re[\varphi_1] (Y^a A_a + \lambda).
\end{align}
As discussed below \cref{eq:Q_C_integral}, the difference between the canonical and Noether fluxes can also be expressed purely in terms of the radiative degrees of freedom. However, the difference between the Noether and stress-energy fluxes does depend on the Coulombic degrees of freedom even when computed over all of \(\scri\), except when $Y^a = 0$ and $\lambda = 0$ (a pure supertranslation).

We stress once more that none of these fluxes can be written as the difference of charges evaluated on cross-sections of null infinity. Thus, on a non-dynamical background spacetime, none of these fluxes can be considered as the change of energy or angular momentum at a particular ``time'' (a cross-section of null infinity), and there is no obvious criterion to decide which of these currents defines the flux of energy or angular momentum.

\section{Einstein-Maxwell theory}
\label{sec:GR-EM}

In this section, we review the symplectic structure at $\scri$ as well as the asymptotic behaviour of asymptotically flat spacetimes in Einstein-Maxwell theory. The reader familiar with this can safely skip to the next section.

\subsection{Symplectic current for Einstein-Maxwell theory}
Following~\cite{burnett-wald}, the Lagrangian for Einstein-Maxwell theory is given by
\begin{equation}
  \df L = \frac{1}{16 \pi} \left(\hat{R} - \hat{F}^2\right) \hat{\df \varepsilon}_4.
\end{equation}
As in the case of Maxwell theory, our analysis is unaffected by adding additional matter sources of compact support or sufficiently fast falloff at null infinity.

A variation of this Lagrangian with respect to the dynamical fields \(\hat\Phi = (\hat g_{ab}, \hat A_a)\) gives (raising and lowering with the background physical metric)
\begin{equation}
  \delta \df L = \left[- \frac{1}{16 \pi} (\hat{G}^{ab} - 8\pi \hat{T}^{ab}) \delta \hat{g}_{ab} + \frac{1}{4\pi} \hat{\nabla}_b \hat{F}^{ba} \delta \hat{A}_a \right] \hat{\df \varepsilon}_4 + d\df\theta(\delta \hat\Phi),
\end{equation}
where \(\hat G_{ab}\) is the Einstein tensor of \(\hat g_{ab}\) and the stress-energy tensor \(\hat T_{ab}\) is the same as in \cref{eq:M_T}, except that the spacetime metric is now also dynamical. The variations with respect to the dynamical fields \( \hat\Phi = (\delta \hat g_{ab}, \delta \hat A_a)\) give the Einstein equations and Maxwell equations, respectively:
\begin{equation}
  \hat{G}_{ab} = 8\pi \hat{T}_{ab} \eqsp \hat{\nabla}_b \hat{F}^{ba} = 0.
\end{equation}
The symplectic potential \(\df\theta\) is given by
\begin{equation}\label{eq:theta-total}\begin{aligned}
  \df\theta(\hat{\Phi}; \delta \hat{\Phi}) &\equiv \hat\varepsilon_{dabc} \hat v^d, \\
    \text{with} \quad
  \hat{v}^a & = \frac{1}{8 \pi} \left(\hat g^{a[b} \hat g^{c]d} \hat \nabla_c \delta \hat g_{bd} - 2 \hat{F}^{ab} \delta \hat{A}_b\right)\, ,
\end{aligned}\end{equation}
where the second term is the symplectic potential of electromagnetism from \cref{eq:theta-EM}. The symplectic current \(\df\omega \defn \delta_1 \df\theta(\delta_2 \hat\Phi) - \delta_2 \df\theta(\delta_1 \hat\Phi) \) is given by the sum of three terms (see Eq.~3.12 of \cite{burnett-wald})\footnote{Note that our expressions \cref{eq:omega-EM,eq:omega_crossterm} differ in appearance from the ones in Eq.~3.12 of \cite{burnett-wald} only because \cite{burnett-wald} uses the perturbed quantity \(\delta \hat{F}^{ab}\) while we prefer to use \(\delta \hat{F}_{ab}\).}
\begin{equation}\label{eq:symp-total}
  \df\omega(\delta_1 \hat{\Phi}; \delta_2 \hat{\Phi}) \equiv \hat\varepsilon_{dabc} \left[\hat{w}_{\rm{GR}}^d (\delta_1 \hat{g}, \delta_2 \hat{g}) + \hat{w}_{\rm{EM}}^d (\delta_1 \hat{A}, \delta_2 \hat{A}) + \hat{w}_\times^d (\delta_1 \hat{\Phi}, \delta_2 \hat{\Phi})\right].
\end{equation}
The first term on the right-hand side of \cref{eq:symp-total} is the same as the symplectic current for vacuum general relativity (see Eqs.~41 and 42 of \cite{WZ}):
\begin{equation}
  \begin{split}
    \hat{w}_{\rm{GR}}^a (\delta_1 \hat{g}, \delta_2 \hat{g}) = \frac{1}{16\pi} \hat{P}^{abcdef} \left[\delta_2 \hat{g}_{bc} \hat{\nabla}_d \delta_1 \hat{g}_{ef} - (1 \lra 2)\right],
  \end{split}
\end{equation}
with
\be\label{eq:P-defn}
    \hat P^{abcdef} = \hat{g}^{ae}\hat{g}^{fb}\hat{g}^{cd} - \tfrac{1}{2}\hat{g}^{ad}\hat{g}^{be}\hat{g}^{fc} - \tfrac{1}{2}\hat{g}^{ab}\hat{g}^{cd}\hat{g}^{ef} - \tfrac{1}{2}\hat{g}^{bc}\hat{g}^{ae}\hat{g}^{fd} + \tfrac{1}{2}\hat{g}^{bc}\hat{g}^{ad}\hat{g}^{ef} \, .
\ee
Similarly, the second term is the symplectic current of electromagnetism from \cref{eq:M_symp_current}:
\begin{equation}\label{eq:omega-EM}
  \hat{w}_{\rm{EM}}^a (\delta_1 \hat{A}, \delta_2 \hat{A}) = -\frac{1}{4\pi} \hat{g}^{ac} \hat{g}^{bd} \lb[ \delta_1 \hat{F}_{cd} \delta_2 \hat{A}_b - (1 \lra 2) \rb] \, ,
\end{equation}
while the third ``cross-term'' is given by
\begin{equation} \label{eq:omega_crossterm}
  \hat w_\times^a (\delta_1 \hat{\Phi}, \delta_2 \hat{\Phi}) = -\frac{1}{4\pi} \left(2 \hat{g}^{c[a} \hat{F}^{b]d} + \tfrac{1}{2} \hat{F}^{ab} \hat{g}^{cd}\right) \delta_2 \hat{A}_b \delta_1 \hat{g}_{cd} - (1 \lra 2).
\end{equation}
This cross-term is unimportant for our analysis, as it vanishes in the limit to $\scri$ for asymptotically flat perturbations.

\subsection{Asymptotic conditions and field equations at \(\scri\)}
\label{sec:GR-EM-scri}

We now review the asymptotic behaviour of Einstein-Maxwell theory near $\scri$.
We use the standard definition of asymptotic flatness (see, for instance \cite{Geroch-asymp}).
The addition of electromagnetic fields does not spoil this definition, since $F_{ab} = \hat{F}_{ab}$ has a smooth extension to $\scri$.

Using the conformal transformation relating the unphysical Ricci tensor \(R_{ab}\) to the physical Ricci tensor \(\hat R_{ab}\) (see Appendix~D of \cite{Wald-book}), the Einstein equation \(\hat G_{ab} = 8 \pi \hat T_{ab}\) can be written as
\begin{equation}\label{eq:S-EE}
  S_{ab} = -2 \Omega^{-1} \nabla_a n_b + \Omega^{-2} n^c n_c g_{ab} + 8\pi \Omega^2 \left(T_{ab} - \tfrac{1}{3} g_{ab} g^{cd} T_{cd}\right) \, ,
\end{equation}
where \(S_{ab}\) and \(T_{ab}\) are given, respectively, by
\be\label{eq:S-T-defn}
    S_{ab} \defn R_{ab} - \tfrac{1}{6} R g_{ab} \eqsp T_{ab} \defn \Omega^{-2}\hat{T}_{ab} \, .
\ee
For Maxwell fields, we have, by \cref{eq:M_T} and the asymptotic conditions in \cref{eq:misc_smooth},
\begin{equation}\label{eq:T-Maxwell}
  T_{ab} = \frac{1}{4\pi} \left(F_{ac} F_b{}^c - \tfrac{1}{4} g_{ab} F^{cd} F_{cd} \right).
\end{equation}
This quantity is smooth at \(\scri\) by the smoothness of $F_{ab}$ and $g_{ab}$.

As before, we assume that the conformal factor is chosen to satisfy the Bondi condition \cref{eq:Bondi-cond,eq:nn-cond}:
\be
    \nabla_a n_b \hateq 0 \eqsp n_a n^a = O(\Omega^2).
\ee

Furthermore, without loss of generality, the conformal factor \(\Omega\) in a neighbourhood of \(\scri\) and the unphysical metric \(g_{ab}\vert_\scri\) at \(\scri\) may be assumed to be universal, i.e., independent of the choice of physical metric \(\hat g_{ab}\) \cite{GW,WZ} (see Appendix~A of \cite{FPS} for details of the argument). Now consider a physical metric perturbation \(\delta \hat g_{ab}\). Since the conformal factor can be chosen universally, we have
\be
    \delta g_{ab} = \Omega^2 \delta \hat g_{ab} \, .
\ee
Given that the unphysical metric \(g_{ab}\vert_\scri\) at \(\scri\) is universal, \(\delta g_{ab} \hateq 0\), and thus there exists a smooth tensor field \(\tau_{ab}\) such that
\be
    \delta g_{ab} = \Omega \tau_{ab} \, .
\ee
Furthermore, imposing the Bondi condition on the perturbations, i.e., \(\delta (\nabla_a n_b) \hateq 0\), we also find (see Eqs.~51--53 of \cite{WZ})
\be
    \tau_{ab} n^b = \Omega \tau_a
\ee
for some smooth \(\tau_a\). Thus, our asymptotic conditions on the metric perturbations imply that
\begin{equation} \label{eq:asymp-flat-conds}
  \tau_{ab} \defn \Omega^{-1} \delta g_{ab} \eqsp \tau_a \defn \Omega^{-1} \tau_{ab} n^b
\end{equation}
are smooth on $\scri$.

For the Maxwell field, we will use the same conditions as in \cref{sec:EM-currents-scri}; that is, \(A_a = \hat A_a\) is smooth at \(\scri\) and satisfies \(n^a A_a \hateq 0\) (\cref{eq:A_bondi}).

\section{Wald-Zoupas charges and fluxes}
\label{sec:WZ}

In this section we derive the charges and fluxes associated with asymptotic symmetries in Einstein-Maxwell theory at null infinity using the Wald-Zoupas prescription. We first review the Wald-Zoupas procedure for obtaining charges and fluxes corresponding to asymptotic symmetries for a general diffeomorphism-covariant theory in \cref{sec:WZ-summ}, and then we apply this prescription to the Einstein-Maxwell case in \cref{sec:WZ-GREM}. We show that the contribution of the Maxwell fields to the Wald-Zoupas flux is given by the Noether current and \emph{not} the stress-energy current. Furthermore, this flux can be determined entirely from the radiative degrees of freedom, and the total flux over all of \(\scri\) acts as a Hamiltonian generator on the radiative phase space.\\

\subsection{Summary of the Wald-Zoupas prescription}
\label{sec:WZ-summ}

The prescription of Wald and Zoupas can be applied to any local and covariant theory. We review below the essential ingredients, emphasizing the subsequent application to Einstein-Maxwell theory.

When the dynamical fields \(\hat\Phi\) satisfy the equations of motion, and \(\delta \hat\Phi\) satisfy the linearized equations of motion, one can show that (see \cite{lee-wald,iyer-wald,KP-bundle})
\begin{equation} \label{eq:omega_symmetry}
  \df \omega(\hat\Phi; \delta \hat\Phi, \delta_{\hat\xi} \hat\Phi) = d \left[\delta \df Q[\hat\xi] - \hat X \cdot \df \theta(\delta \hat\Phi)\right]
\end{equation}
for all symmetries \(\hat\xi\), where the \(2\)-form \(\df Q[\hat\xi]\) is the \emph{Noether charge} associated with the symmetry \(\hat\xi\). In Einstein-Maxwell theory, \(\df Q[\hat\xi]\) is given by
\be
    \df Q[\hat\xi] \equiv - \frac{1}{16\pi} \hat\varepsilon_{cd ab} \hat\nabla^c \hat X^d - \frac{1} {8\pi} \hat\varepsilon_{cd ab} \hat F^{cd} (\hat X^e \hat A_e + \hat\lambda) \, .
\ee
The first term above is the Noether charge associated with the vector field \(\hat X^a\) in vacuum general relativity (Eq.~44 \cite{WZ}), and the second term is the Noether charge for electromagnetism given in \cref{eq:Q-N-EM}.

Now we consider \cref{eq:omega_symmetry} at \(\scri\), rewritten in terms of the unphysical fields which are smooth at \(\scri\). Using \cref{eq:asymp-flat-conds,eq:misc_smooth}, it can be verified that the symplectic current \(\df\omega\) (\cref{eq:symp-total}) has a limit to \(\scri\). Thus, from this point onward, we work with the fields and symmetries in the unphysical spacetime. Now, consider a spacelike surface $\Sigma$ which intersects \(\scri\) at some cross-section \(S\). Integrating \cref{eq:omega_symmetry} over $\Sigma$, we then find
\begin{equation}\label{eq:omega-dQ}
  \int_\Sigma \df\omega(\Phi; \delta\Phi, \delta_\xi \Phi) = \int_S \big(\delta \df Q[\xi] - X \cdot \df \theta(\delta \Phi)\big).
\end{equation}
Since \(\df\omega\) admits a limit to \(\scri\), the integral on the left-hand side of \cref{eq:omega-dQ} is always finite. However, the \(2\)-form integrand on the right-hand side need not have a finite limit to \(\scri\) in general. Thus, the integral on the right-hand side of \cref{eq:omega-dQ} should be understood as being defined by first integrating over some $2$-sphere in $\Sigma$ and then taking the limit of this \(2\)-sphere to $S$ \cite{WZ}. This final limiting integral is independent of the way in which the limits are taken since \(d\df\omega(\Phi; \delta\Phi, \delta_\xi \Phi) = 0\).

From the above identity, it would be natural to define a charge associated with the asymptotic symmetry \(\xi\) at \(S\) as a function \(Q[\xi; S]\) in the phase space of the theory such that
\be\label{eq:Q-naive}
    \delta Q[\xi; S] = \int_S \big(\delta \df Q[\xi] - X \cdot \df \theta(\delta \Phi)\big)
\ee
for all perturbations \(\delta\Phi\). However, in general, no such charge exists, since the right-hand side is not integrable in phase space, i.e., cannot be written as $\delta (\rm{something})$ for \emph{all} perturbations. To see this, suppose that the charge defined in \cref{eq:Q-naive} does exist. Then, one must have \((\delta_1 \delta_2 - \delta_2 \delta_1) Q[\xi; S] = 0\) for \emph{all} backgrounds \(\Phi\) and \emph{all} perturbations \(\delta_1\Phi, \delta_2\Phi\) (satisfying the corresponding equations of motion). However, it is straightforward to compute that
\be\label{eq:Q-integrablity}
    (\delta_1 \delta_2 - \delta_2 \delta_1) Q[\xi; S] = - \int_S X \cdot \df\omega(\Phi; \delta_1\Phi, \delta_2 \Phi) \, .
\ee
Thus, a charge defined by \cref{eq:Q-naive} will exist if the right-hand side of the above equation vanishes. This is the case in Einstein-Maxwell theory if \(X^a \hateq 0\), i.e., for a pure asymptotic Maxwell gauge symmetry, or if \(X^a\) is tangent to \(S\). However in general, the right-hand side is non-vanishing and one cannot define any charge \(Q[\xi;S]\) using \cref{eq:Q-naive}.

This obstruction is resolved by the rather general prescription of Wald and Zoupas \cite{WZ}. Their procedure for defining integrable charges associated with asymptotic symmetries can be summarized as follows: let $\df \Theta(\delta \Phi)$ be a symplectic potential for the pullback of $\df \omega$ to $\scri$, i.e.,
\begin{equation}\label{eq:Theta-defn}
  \pb{\df \omega} (\delta_1\Phi, \delta_2 \Phi) = \delta_1 \df \Theta(\delta_2 \Phi) - \delta_2 \df \Theta(\delta_1 \Phi)
\end{equation}
for all backgrounds and all perturbations (with suitable asymptotic conditions and equations of motion imposed). Following \cite{WZ}, we require that
\begin{enumerate}
    \item \(\df\Theta\) be locally and covariantly constructed out of the dynamical fields \(\Phi\), \(\delta \Phi\), and finitely many of their derivatives, along with any fields in the ``universal background structure'' present at \(\scri\).
    \item \(\df\Theta\) be independent of any arbitrary choices made in specifying the background structure; i.e., \(\df\Theta\) is conformally invariant as well as invariant under Maxwell gauge transformations on \(\scri\) for Einstein-Maxwell theory. We also require that \(\df\Theta\) be independent of the choice of the auxiliary normal \(l^a\) and the corresponding \(q^{ab}\) used in our computations.
    \item  $\df \Theta(\delta \Phi) = 0$ for \emph{any} stationary background solution \(\Phi\) and for \emph{all} (not necessarily stationary) perturbations $\delta \Phi$.
\end{enumerate}

If such a symplectic potential \(\df\Theta\) can be found, define \(\mc Q[\xi; S]\) to be a function on the phase space at \(\scri\) by\footnote{Note that the first of these two integrals is defined by the limiting procedure described below \cref{eq:omega-dQ}, whereas the second is an ordinary integral, as $\df \Theta$ is defined directly on $\scri$.}
\begin{equation}\label{eq:WZ-defn}
  \delta \mc Q[\xi; S] \defn \int_S \big(\delta \df Q[\xi] - X \cdot \df \theta(\delta \Phi)\big) + \int_S X \cdot \df \Theta(\delta \Phi) \, .
\end{equation}
It is easily checked (using \cref{eq:Q-naive,eq:Q-integrablity,eq:Theta-defn}) that this expression is integrable in phase space, i.e., $(\delta_1 \delta_2 - \delta_2  \delta_1) \mc Q[\xi; S] = 0$. Together with some choice of reference solution \(\Phi_0\) on which \(\mc Q[\xi;S] = 0\) for all asymptotic symmetries \(\xi\) and all cross-sections \(S\), \cref{eq:WZ-defn} defines the \emph{Wald-Zoupas charge} \(\mc Q[\xi;S]\) associated with the asymptotic symmetry \(\xi\) at \(S\).

 The flux of the perturbed Wald-Zoupas charge is given by (see also Eqs.~28 and 29 of \cite{WZ})
\be\label{eq:delta-F1}
    \delta \mc F[\xi;\Delta\scri] \defn \delta \mc Q[\xi; S_2] - \delta \mc Q[\xi; S_1] = - \int_{\Delta \scri} \big[ \pb{\df\omega}(\delta\Phi, \delta_\xi \Phi) + d [X \cdot \df\Theta(\delta\Phi)] \big].
\ee
The last term of this equation can also be written as
\be\begin{aligned}
    d [X \cdot \df\Theta(\delta\Phi)] & = \Lie_X \df\Theta(\delta\Phi) \\
    & = \delta_\xi \df\Theta(\delta\Phi) \\
    & = - \pb{\df\omega}(\delta\Phi, \delta_\xi \Phi) + \delta \df\Theta(\delta_\xi\Phi ) \, ,
\end{aligned}\ee
where in the second line we have used the criteria that \(\df\Theta\) is a local and covariant functional on \(\scri\) and that it is invariant under Maxwell gauge transformations,\footnote{In the principal bundle language, this means \(\df\Theta\) is a gauge-invariant and \emph{horizontal} \(3\)-form on the bundle.} while the third line follows from the definition of \(\df\Theta\) as a symplectic potential for \(\pb{\df\omega}\) (\cref{eq:Theta-defn}). Thus, the flux of the perturbed Wald-Zoupas charge is
\begin{equation}
  \delta \mc F[\xi;\Delta\scri] = - \int_{\Delta \scri} \delta \df \Theta(\delta_\xi \Phi).
\end{equation}

To get the unperturbed charge and flux from the perturbed ones we have to choose a reference solution $\Phi_0$ on which the charges are required to vanish. Since the \(\df\Theta(\delta\Phi)\) is required to vanish on stationary backgrounds we choose the reference solution \(\Phi_0\) to also be stationary. For our concrete case of Einstein-Maxwell theory, we will pick \(\Phi_0\) to be Minkowski spacetime. Then, the flux of the Wald-Zoupas charge is simply
\begin{equation} \label{eq:general-flux}
  \mc F[\xi; \Delta\scri] = \mc Q[\xi; S_2] - \mc Q[\xi; S_1] = -\int_{\Delta \scri} \df \Theta(\delta_\xi \Phi).
\end{equation}

Note that from \cref{eq:delta-F1} we also have
\begin{equation} \label{eq:hamiltonian}
  \delta \mc F[\xi; \Delta\scri] = - \int_{\Delta \scri} \pb{\df \omega} (\delta \Phi, \delta_\xi \Phi) + \int_{S_2}  X \cdot \df \Theta(\delta \Phi) - \int_{S_1} X \cdot \df \Theta(\delta \Phi).
\end{equation}
If the boundary terms on \(S_2\) and \(S_1\) vanish for all backgrounds \(\Phi\) and all perturbations \(\delta\Phi\) then  $\mc F[\xi; \Delta\scri]$ also defines a \emph{Hamiltonian generator} (relative to the symplectic current \(\pb{\df\omega}\)) on the radiative phase space on $\Delta \scri$ corresponding to the symmetry \(\xi\). For general field configurations, these boundary terms do not vanish on finite cross-sections of \(\scri\). However, we will show below in Einstein-Maxwell theory that when \(\Delta\scri\) is taken to be all of null infinity, appropriate boundary conditions at timelike and spacelike infinity (i.e, as \(|u| \to \infty \)) ensure that these boundary terms indeed vanish for our choice of \(\df\Theta\). Thus, our fluxes define the Hamiltonian generators for Einstein-Maxwell theory on the phase space on all of \(\scri\).

\begin{remark}[Ambiguities in the Wald-Zoupas prescription]
\label{rem:amb}
For a given theory, the Wald-Zoupas prescription is not unambiguously defined. For a given Lagrangian \(\df L\), the symplectic potential \(\df\theta\) is ambiguous up to the redefinition
\be
    \df\theta(\delta \hat\Phi) \mapsto \df\theta(\delta \hat\Phi) + d \df Y(\delta \hat\Phi)
\ee
where \(\df Y(\delta \hat\Phi)\) is a local and covariant \(2\)-form which is a linear functional of the perturbations \(\delta \hat\Phi\) and finitely many of its derivatives. This changes the symplectic current by
\be
    \df\omega(\delta_1 \hat\Phi, \delta_2 \hat\Phi) \mapsto \df\omega(\delta_1 \hat\Phi, \delta_2 \hat\Phi) + d \lb[\delta_1 \df Y(\delta_2 \hat\Phi) - \delta_2 \df Y(\delta_1 \hat\Phi)\rb] \,.
\ee 
Note that the addition of a boundary term to the Lagrangian does not affect the symplectic form. Even with a fixed choice of the symplectic current, the symplectic potential \(\df\Theta(\delta \Phi)\) defined on null infinity (\cref{eq:Theta-defn}) is ambiguous up to
\be
    \df\Theta(\delta\Phi) \mapsto \df\Theta(\delta\Phi) + \delta \df W(\Phi)
\ee
where \(\df W\) is a local and covariant \(3\)-form on \(\scri\). These ambiguities then also lead to ambiguities in the Wald-Zoupas prescription for the charges and fluxes on null infinity. It was argued by Wald and Zoupas that these ambiguities do not affect their prescription in vacuum GR (see footnote~18 and the arguments below Eq.~73 in \cite{WZ}). We hope that similar arguments can also be made for Einstein-Maxwell theory, but we do not analyze these ambiguities in detail.
\end{remark}

\subsection{Computation of the Wald-Zoupas charges and fluxes at null infinity for Einstein-Maxwell theory}
\label{sec:WZ-GREM}

We now apply the above described prescription of Wald and Zoupas to Einstein-Maxwell theory and compute the charges and fluxes at \(\scri\). Since our main focus is on the contribution of the Maxwell fields to the charges and fluxes, we will borrow the analysis of Wald and Zoupas \cite{WZ} for the contribution of the gravitational field.

 First, we compute the pullback \(\pb{\df\omega}\) to \(\scri\) of the symplectic current in \cref{eq:symp-total}. Using the asymptotic conditions \cref{eq:asymp-flat-conds,eq:misc_smooth}, it can be checked that the contribution from the cross-term given by \(- \Omega^{-4} n_a \hat w_{\times}^a\) (\cref{eq:omega_crossterm}) vanishes in the limit to \(\scri\). The contribution from the Maxwell fields is easily computed to be
\begin{equation} \label{eq:omega_EM}
  \pb{\df \omega_{\rm{EM}}} (\delta_1 A, \delta_2 A) \hateq - \Omega^{-4} n_a \hat w_{\rm EM}^a ~\df\varepsilon_3 = -\frac{1}{4 \pi} \left[\delta_1 \mc E^a \delta_2 A_a - \delta_2 \mc E^a \delta_1 A_a\right] \df \varepsilon_3.
\end{equation}
The contribution from the metric perturbations is the most tedious to compute. However, since the $T_{ab}$ for Maxwell fields is smooth on $\scri$, the terms proportional to the stress-energy tensor in \cref{eq:S-EE} vanish at \(\scri\), and the computation of \cite{WZ} carries over unchanged. We therefore find (see Eq.~72 of \cite{WZ})\footnote{As mentioned before, one can consider additional sources with compact support or sufficient falloff at \(\scri\) without affecting this analysis.}
\begin{equation} \label{eq:omega_GR}
  \pb{\df \omega_{\rm GR}} (\delta_1 g, \delta_2 g) \hateq - \Omega^{-4} n_a \hat w_{\rm GR}^a ~\df\varepsilon_3 = -\frac{1}{32 \pi} \left[\delta_1 N_{ab} \tau_2^{ab} - \delta_2 N_{ab} \tau_1^{ab} \right] \df \varepsilon_3.
\end{equation}
Here \(N_{ab}\) is the \emph{News tensor} on \(\scri\) defined by
\be
    N_{ab} \defn \pb{S_{ab}} - \rho_{ab},
\ee
where \(\pb{S_{ab}}\) is the pullback to \(\scri\) of \(S_{ab}\) and \(\rho_{ab}\) is the unique symmetric tensor field on \(\scri\) constructed from the universal structure at \(\scri\) in Theorem~5 of \cite{Geroch-asymp}. The News tensor also satisfies the properties
\be\label{eq:News-conds}
    N_{ab} n^b \hateq 0 \eqsp N_{ab} q^{ab} \hateq 0.
\ee
Thus, the pullback to \(\scri\) of the symplectic current of Einstein-Maxwell theory is given by
\be\label{eq:omega-pb}
    \pb{\df\omega} = -\frac{1}{32 \pi} \left[\delta_1 N_{ab} \tau_2^{ab} - \delta_2 N_{ab} \tau_1^{ab} \right] \df \varepsilon_3 -\frac{1}{4 \pi} \left[\delta_1 \mc E^a \delta_2 A_a - \delta_2 \mc E^a \delta_1 A_a\right] \df \varepsilon_3 \, .
\ee

Note that \(\pb{\df\omega}\) is determined completely by the (perturbed) radiative degrees of freedom. For the Maxwell fields, it is clear that only the perturbations of \(\pb{A_a}\) and \(\mc E_a = - \Lie_n \pb{A_a}\) contribute. For the gravitational fields, the argument is more involved. Consider the \emph{asymptotic shear} of the cross-sections of \(\scri\) defined by
\be\label{eq:sigma-defn}
    \sigma_{ab} \defn (q_a{}^c q_b{}^d - \tfrac{1}{2} q_{ab} q^{cd} ) \nabla_c l_d \, ,
\ee
which is related to the News tensor through
\be
    N_{ab} = 2 \Lie_n \sigma_{ab} \, .
\ee
Using the asymptotic conditions \cref{eq:asymp-flat-conds}, the perturbation of the shear generated by the metric perturbation \(\delta g_{ab}\) (with fixed \(l_a\), since \(l_a\) can be chosen independently of the spacetime) can be computed to be
\be\label{eq:sigma-tau}
    \delta \sigma_{ab} \hateq - \tfrac{1}{2} (q_a{}^c q_b{}^d - \tfrac{1}{2} q_{ab} q^{cd} ) \tau_{cd};
\ee
that is, \(\delta \sigma_{ab}\) is given by the trace-free part of \(\tau_{ab}\) on the cross-sections. Due to the conditions \cref{eq:News-conds} and \(\tau_{ab}n^b \hateq 0\) (from \cref{eq:asymp-flat-conds}), it is clear that only this trace-free part of \(\tau_{ab}\) --- equivalently, \(\delta \sigma_{ab}\) --- contributes to the pullback of the symplectic current. Furthermore, from the analysis of Ashtekar and Streubel \cite{Ash-Str}, \(\delta\sigma_{ab}\) is equivalent to the perturbation in the equivalence class of derivatives \(\{D_a\}\) defined on \(\scri\), which are the radiative degrees of freedom in vacuum GR. Thus, \(\pb{\df\omega}\) is completely determined by the perturbed radiative degrees of freedom in Einstein-Maxwell theory. The integral of this symplectic current over all of \(\scri\) (when appropriate falloff conditions are satisfied toward \(i^0\) and \(i^+\); see \cref{eq:E-N-conds}) reproduces the symplectic form on the radiative phase space at null infinity used by Ashtekar and Streubel \cite{Ash-Str}.\\

To apply the Wald-Zoupas prescription, we need to find a \(3\)-form symplectic potential \(\df\Theta(\delta\Phi)\) for \(\pb{\df\omega}\) given in \cref{eq:omega-pb}. We choose the following (see \cref{rem:amb} for the ambiguities in the choice of \(\df\Theta\)):
\begin{equation}\label{eq:Theta-guess}\begin{aligned}
  & \df \Theta(\delta\Phi)  = \df \Theta_{\rm{GR}} (\delta g) + \df \Theta_{\rm{EM}} (\delta A),\\
   & \text{where} \quad
    \df \Theta_{\rm{GR}} (\delta g)  = -\frac{1}{32 \pi} N_{ab} \tau^{ab} ~\df\varepsilon_3 \\
   & \quad \quad \quad  \; \; \df \Theta_{\rm{EM}} (\delta A) = -\frac{1}{4 \pi} \mc E^a \delta A_a ~\df\varepsilon_3.
\end{aligned}\end{equation}
Note that \(\df \Theta_{\rm{GR}} (\delta g)\) is the symplectic potential for vacuum GR given in Eq.~73 of \cite{WZ}. The above choice of \(\df\Theta\) satisfies all the requirements listed below \cref{eq:Theta-defn}:
\begin{enumerate}
    \item The \(\df\Theta\) in \cref{eq:Theta-guess} is indeed a local and covariant functional of the background fields \(\Phi\) and the perturbed fields \(\delta\Phi\) (see also footnote~20 of \cite{WZ} for an explanation of the locality of the News tensor).
    \item It is also invariant under conformal transformations and Maxwell gauge transformations,\footnote{Note that \(\delta A_a\) is gauge invariant, since the gauge transformations are independent of the dynamical fields.} and the choice of the auxiliary null normal \(l^a\) and the ``inverse metric'' \(q^{ab}\).
    \item As we show in \cref{sec:stationary}, for stationary solutions of Einstein-Maxwell theory we have \(\mc E_a = 0\) and \(N_{ab} = 0\) on \(\scri\), and thus \(\df\Theta(\Phi;\delta\Phi)\), as defined above, vanishes for all perturbations \(\delta\Phi\) whenever the background \(\Phi\) is a stationary solution of the Einstein-Maxwell equations.
\end{enumerate}

Having chosen a \(\df\Theta\) as in \cref{eq:Theta-guess} the Wald-Zoupas flux \(\mc F[\xi; \Delta\scri]\) associated with an asymptotic symmetry \(\xi\) is determined by \cref{eq:general-flux}. We now want to find the corresponding Wald-Zoupas charge \(\mc Q[\xi;S]\) on any cross-section \(S\) of \(\scri\). Note that the Wald-Zoupas charge is determined by \cref{eq:WZ-defn}, along with the requirement that it vanish on some stationary reference solution \(\Phi_0\) which we take to be Minkowski spacetime. Although the right-hand side of \cref{eq:WZ-defn} can be directly computed, it is not very useful to find an expression for \(\mc Q[\xi;S]\). We instead proceed in the following manner: let the Wald-Zoupas charge be given by
\begin{equation}\label{eq:WZ-total}
  \mc Q[\xi; S] = \mc Q_{\rm{GR}}[X; S] + \mc Q_{\rm{EM}}[\xi; S],
\end{equation}
where $\mc Q_{\rm{GR}}[X; S]$ is the expression for the charge in vacuum GR (see \cref{eq:Q-GR-defn}) and \(\mc Q_{\rm{EM}}[\xi; S]\) is the (as yet undetermined) contribution due to the Maxwell fields. As we will show below, in the presence of Maxwell fields, \(\mc Q_{\rm{GR}}[X; S]\) by itself does not satisfy \cref{eq:general-flux} with \(\df\Theta\) as in \cref{eq:Theta-guess}; that is, \(\mc Q_{\rm{GR}}[X; S]\) is not the full Wald-Zoupas charge for Einstein-Maxwell theory. Then, we will define the Maxwell contribution \(\mc Q_{\rm{EM}}[\xi; S]\) so that the total charge \cref{eq:WZ-total} does satisfy \cref{eq:general-flux,eq:Theta-guess}, and \(\mc Q_{\rm{EM}}[\xi; S]\) vanishes in the absence of the electromagnetic field.

In vacuum GR, the Wald-Zoupas charge for a BMS vector field \(X^a\) can be written as follows. With our assumptions on the asymptotic conditions on the fields, it follows that \(C_{abcd} \hateq 0\) (see Theorem~11 of \cite{Geroch-asymp}), and thus \(\Omega^{-1}C_{abcd}\) is smooth at \(\scri\). Then \(\mc Q_{\rm GR}\) is given by
\be\label{eq:Q-GR-defn}
    \mc Q_{\rm{GR}}[\xi; S] = \frac{1}{8\pi} \int_S \df\varepsilon_2 \left[ - X^a (\Omega^{-1}C_{abcd}) l^b l^c n^d + \tfrac{1}{2} \beta \sigma^{ab} N_{ab} + Y^a \sigma_{ab} \ms D_c \sigma^{bc} - \tfrac{1}{4} \sigma_{ab} \sigma^{ab} \ms D_c Y^c \right],
\ee
where we have decomposed \(X^a \hateq \beta n^a + Y^a\), with \(Y^a\) tangent to the cross-sections of the chosen foliation (see \cref{eq:BMS-decomp}). The tensor \(\sigma_{ab}\) is the asymptotic shear of the cross-sections defined in \cref{eq:sigma-defn}.

For \emph{vacuum} GR, the charge expression \cref{eq:Q-GR-defn} coincides with the charges defined by Wald and Zoupas \cite{WZ}. Showing this explicitly is a long and tedious computation, but we argue as follows. For supertranslations, \cref{eq:Q-GR-defn} is the same as the supermomentum defined by Geroch \cite{Geroch-asymp}, which is equal to the Wald-Zoupas charge (see Eq.~98 of \cite{WZ}). For asymptotic Lorentz symmetries, it was shown in \cite{WZ} that the Wald-Zoupas charge is given by the ``linkage'' charge\footnote{Note that for general supertranslations the ``linkage'' charges and fluxes do not equal the ones obtained from Hamiltonian methods \cite{Ash-Str} or from the Wald-Zoupas prescription; see \cite{AW-linkage}.} found by Geroch and Winicour \cite{GW}, which, in turn, coincides with the above expression as shown by Winicour \cite{Winicour}. The expression \cref{eq:Q-GR-defn} is also equal to the charge found in \cite{AK}, when the conformal factor is additionally chosen away from \(\scri\) to make the vector field \(l^a\) expansion-free. It is also equal to the expression computed using Bondi coordinates (see, for instance, Eq.~35 of \cite{flanagan-nichols}).

In vacuum GR, the flux of the charge \cref{eq:Q-GR-defn} is given by \cref{eq:general-flux}, with $\df \Theta_{\rm{GR}} (\Lie_X g)$ on the right-hand side. However, in the presence of Maxwell fields one gets an additional contribution to the flux of this charge through the asymptotic stress-energy tensor \(T_{ab}\). This additional contribution arises through the \(\Lie_n\) of the Weyl tensor term, and using the Bianchi identity at \(\scri\) we get\footnote{In the Newman-Penrose notation, the Weyl tensor terms appearing in \cref{eq:Q-GR-defn} are \(\Re[\psi_2]\) and \(\psi_1\). Their derivatives on \(\scri\) along \(n^a\) are determined by the Bianchi identities given in Eqs.~9.10.5 and 9.10.6 of \cite{PR2}.}
\begin{equation}\label{eq:Q-GR-flux}
  \mc Q_{\rm{GR}}[X; S_2] - \mc Q_{\rm{GR}}[X; S_1] = - \int_{\Delta \scri} \left[\df \Theta_{\rm{GR}} (\Lie_X g) + T_{ab} n^a X^b \df \varepsilon_3\right].
\end{equation}
If one takes \(\mc Q_{\rm GR}\) as the definition of the charges associated with the BMS symmetries, then the Maxwell fields contribute to the flux only through the asymptotic stress-energy tensor \(T_{ab}\) (see also Appendix~C of \cite{flanagan-nichols}). As argued in \cref{sec:EM-currents-scri} and in \cite{ab1,ab2}, for Lorentz symmetries this contribution to the flux is \emph{not} purely radiative and depends on the Coulombic part \(\Re[\varphi_1]\) of the Maxwell field. However, in the presence of Maxwell fields at \(\scri\), the usual expression \cref{eq:Q-GR-defn} cannot be the full Wald-Zoupas charge of the theory, as it does not satisfy \cref{eq:general-flux} with the full \(\df\Theta\) in \cref{eq:Theta-guess}, which includes the Maxwell contribution $\df \Theta_{\rm{EM}} (\delta_\xi A)$.

Our goal now is to define the Maxwell contribution \(\mc Q_{\rm EM}\) to the Wald-Zoupas charge such that \(\mc Q_{\rm GR} + \mc Q_{\rm EM}\) satisfies \cref{eq:general-flux} with the full \(\df\Theta\) in \cref{eq:Theta-guess}. From \cref{eq:Theta-guess}, we have for $\df \Theta_{\rm{EM}} (\delta_\xi A)$
\be
    \int_{\Delta \scri} \df \Theta_{\rm{EM}} (\delta_\xi A) =  -\frac{1}{4\pi}\int_{\Delta\scri} \df\varepsilon_3~ q^{ab} \mc E_a (\Lie_X A_b + \ms D_b \lambda).
\ee
This is precisely the flux \(\mc F_{\rm N}[\xi; \Delta\scri]\) of the Noether current of Maxwell theory \cref{eq:flux-N}. This relation arises because, due to our asymptotic conditions, \(\df\Theta_{\rm EM} (\delta A) \hateq \pb{\df\theta}_{\rm EM} (\delta A)\), where the right-hand side is the pullback of the symplectic potential of electromagnetism on a non-dynamical background given in \cref{eq:theta-EM}. It also follows that \(\pb{\df\eta[\xi]} \hateq 0\) (see \cref{eq:eta-defn}), and thus $\df\Theta_{\rm EM}(\delta_\xi A)$ is simply the pullback of the Noether current \(\df J_{\rm N}[\xi]\) for Maxwell theory. Thus, the contribution of the Maxwell field to the flux of the Wald-Zoupas charge is, in fact, the Noether current and not the stress-energy current. This flux contribution is  the same as the one obtained by Ashtekar and Streubel in Eq.~2.18 of \cite{Ash-Str}. However, there the boundary term containing the Coulombic contribution \(\Re[\varphi_1]\) was dropped when converting to the stress-energy expression in Eq.~2.19 of \cite{Ash-Str}. This is valid in their context, as they considered only source-free solutions on Minkowski spacetime (so that  \(\Re[\varphi_1]\) necessarily vanishes); for the more general scenario we are interested in, this boundary term is important and differentiates the Noether and stress-energy current.

From the previous computations, we can relate this Maxwell contribution to the Wald-Zoupas flux to the stress-energy tensor using \cref{eq:F_N,eq:Q_N_value} to get
\be\label{eq:Q-EM-flux}
    \mc Q_{\rm{EM}}[\xi; S_2] - \mc Q_{\rm{EM}}[\xi; S_1] = - \int_{\Delta \scri} \left[\df \Theta_{\rm{EM}} (\delta_\xi A) - T_{ab} n^a X^b \df \varepsilon_3\right],
\ee
where we have defined
\begin{equation}\label{eq:Q-EM-defn}
  \mc Q_{\rm{EM}}[\xi; S] \defn \frac{1}{2\pi} \int_S\df \varepsilon_2 \Re[\varphi_1] (\lambda + X^a A_a),
\end{equation}
which is essentially \cref{eq:Q_N_value} and the integral of the Maxwell Noether charge \cref{eq:Q-N-EM} on the cross-section \(S\). Consequently, from \cref{eq:Q-GR-flux,eq:Q-EM-flux}, it follows that \(\mc Q = \mc Q_{\rm GR} + \mc Q_{\rm EM}\) satisfies
\begin{equation}
  \mc F[\xi; \Delta\scri] = - \int_{\Delta \scri} \df \Theta(\delta_\xi \Phi) = \mc Q[\xi; S_2] - \mc Q[\xi; S_1] \, .
\end{equation}
The Maxwell contribution \(\mc Q_{\rm{EM}}[\xi; S] = 0\) when the Maxwell field \(F_{ab}\) vanishes, and since \(\mc Q_{\rm{GR}}[\xi; S] = 0\) in Minkowski spacetime, the full Wald-Zoupas charge \(\mc Q[\xi; S]\) also vanishes in Minkowski spacetime. 

In sum, the Wald-Zoupas charge for Einstein-Maxwell theory is
\begin{equation}
  \mc Q[\xi; S] = \mc Q_{\rm{GR}}[X; S] + \mc Q_{\rm{EM}}[\xi; S]
\end{equation}
with the individual terms given by \cref{eq:Q-GR-defn,eq:Q-EM-defn}, respectively. The fluxes of the individual terms \(\mc Q_{\rm GR}\) and \(\mc Q_{\rm EM}\) depend on the stress-energy and \emph{cannot} be determined purely from the radiative modes at null infinity. However, from \cref{eq:Q-GR-flux,eq:Q-EM-flux}, these contributions cancel exactly, and so the flux of the full Wald-Zoupas charge \(\mc Q\) can be determined from the radiative modes alone.\\

As mentioned above, the flux $\mc F[\xi;\scri]$ is a Hamiltonian generator on the full radiative phase space of $\scri$, corresponding to the symmetry \(\xi\). Along \(\scri\), as \(u \to \pm \infty\), we have
\be\label{eq:E-N-conds}
    N_{ab} = O(1/|u|^{1+\epsilon} ) \eqsp \mc E_a = O(1/|u|^{1+\epsilon} )
\ee
for some \(\epsilon > 0\), while \(\tau_{ab}\) and \(\delta A_a\) have finite limits as \(u \to \pm \infty\). Note that these conditions are preserved by the asymptotic symmetries. Furthermore, they also ensure that the integral over all of \(\scri\) of the pullback of the symplectic current (\cref{eq:omega-pb}) is finite so that we have a well-defined symplectic form on the radiative phase space on \(\scri\). Since \(X^a\) grows at most linearly in \(u\), from \cref{eq:Theta-guess} we have that
\be
    \lim_{u \to \pm\infty} X \cdot \df\Theta(\delta \Phi) = 0,
\ee
and from \cref{eq:hamiltonian}
\be
    \delta \mc F[\xi; \scri] = -\int_{\scri} \pb{\df \omega} (\delta \Phi, \delta_\xi \Phi),
\ee
for all perturbations \(\delta \Phi\) and all backgrounds \(\Phi\). Thus, the Wald-Zoupas flux acts as a Hamiltonian generator of the corresponding symmetry on the radiative phase space of Einstein-Maxwell theory on all of \(\scri\).\footnote{ 
If one instead defines the flux associated with a BMS symmetry by the right-hand side of \cref{eq:Q-GR-defn}, then such a flux is \emph{not} a Hamiltonian generator in Einstein-Maxwell theory.
}\\

There are several interesting consequences of this result.

First, let us consider the behaviour of the Wald-Zoupas charges under a Maxwell gauge transformation \(A_a \mapsto A_a + \nabla_a \Lambda\) with \(\Lie_n \Lambda \hateq 0\), so that \(n^a A_a \hateq 0\) (\cref{eq:A_bondi}) is preserved. The gravitational contribution \(\mc Q_{\rm GR}\) is, of course, unaffected by this transformation. Similarly, the electromagnetic contribution \(\mc Q_{\rm EM}\) (\cref{eq:Q-EM-defn}) is invariant whenever the asymptotic symmetry \(\xi\) is either a pure Maxwell symmetry \(\xi = (X^a = 0, \lambda)\) or a pure supertranslation \(\xi = (X^a = f n^a, \lambda)\). However, the charge contribution \(\mc Q_{\rm EM}[Y; S]\) for a ``pure Lorentz symmetry'' transforms non-trivially:
\be\label{eq:Q-lor-transform}
    \mc Q_{\rm EM}[Y; S] \mapsto \mc Q_{\rm EM}[Y; S] + \frac{1}{2\pi} \int_S\df \varepsilon_2~ \Re[\varphi_1] \Lie_Y \Lambda \, .
\ee
The second term on the right-hand side is the charge \(\mc Q_{\rm EM} [\Lie_Y \Lambda; S]\) of a pure Maxwell symmetry \(\Lie_Y \Lambda\). Thus, under Maxwell gauge transformation, the electromagnetic contribution to the charge of a Lorentz symmetry shifts by the charge of a pure Maxwell symmetry. This is due to the fact that the action of a ``pure Lorentz symmetry'' \(\xi = (X^a = Y^a, \lambda = 0)\) is not well-defined independently of the choice of gauge for \(A_a\).
This is similar to the transformation of the Lorentz charges under a supertranslation, and it essentially arises due to the fact that the asymptotic symmetry algebra is a semidirect sum of the BMS algebra with the Lie ideal of Maxwell transformations. In the usual BMS algebra for vacuum GR, there is no unique Lorentz subalgebra but instead infinitely many Lorentz subalgebras which are related to each other by supertranslations. Similarly, in Einstein-Maxwell theory, there is no unique action of the Lorentz algebra on the vector potential \(A_a\) at \(\scri\) but infinitely many such actions of the Lorentz algebra which are all related by the asymptotic Maxwell symmetries. Note, however, that taking into account the change of the representation of \(\xi\) in terms of \(X^a\) and \(\lambda\), the charge $\mc Q_{\rm EM}$ is invariant under gauge transformations as follows from \cref{eq:inv}. Essentially, under \(A_a \mapsto A_a + \nabla_a \Lambda\), a ``pure Lorentz symmetry'' is not invariant but transforms as
\be
    (Y^a , \lambda = 0 ) \mapsto (Y^a, -\Lie_Y \Lambda).
\ee
The transformation of the ``pure Lorentz''  charge \cref{eq:Q-lor-transform} is exactly compensated by the transformation of the ``pure Lorentz'' symmetry used to compute the charge.\\

The gravitational fields do not contribute to the Wald-Zoupas charge of a pure Maxwell symmetry \(\xi = (X^a = 0, \lambda)\), which is given by
\be
    \mc Q[\lambda; \Delta\scri] = \mc Q_{\rm{EM}}[\lambda; S] \defn \frac{1}{2\pi} \int_S\df \varepsilon_2 \Re[\varphi_1] ~ \lambda,
\ee
with the flux
\be
    \mc F[\lambda; \Delta\scri] = \frac{1}{4\pi}\int_{\Delta\scri} \df\varepsilon_3~ q^{ab} \mc E_a \ms D_b \lambda.
\ee
For \(\lambda = \text{constant}\), the flux vanishes across any region \(\Delta\scri\), and the charge is proportional to the total conserved Coulomb charge. For a general \(\lambda\) (that is, $\lambda$ is a function on \(\bb S^2\)) this charge is the ``soft charge'' of the Maxwell fields (see~\cite{ab1, asymp-quant}, for example).

Next, consider the charge associated with a supertranslation \(\xi = (X^a \hateq f n^a, \lambda=0)\). Then, the electromagnetic contribution \(\mc Q_{\rm EM}{[f n; S]}\) to the charge vanishes since \(n^a A_a \hateq 0\) and the supermomentum charge is given by the same expression as in vacuum GR. Similarly, from \cref{eq:Q-EM-flux} the Maxwell contribution to the flux of supermomentum is also
\be
    - \int_{\Delta \scri} \df \Theta_{\rm{EM}} (\delta_\xi A) = - \int_{\Delta \scri}\df \varepsilon_3~  f T_{ab} n^a n^b = - \frac{1}{4\pi} \int_{\Delta \scri}\df \varepsilon_3~  f \mc E_a \mc E^a .
\ee
Thus, the electromagnetic fields do not contribute to the supermomentum charge and contribute to the supermomentum flux only through the asymptotic stress-energy tensor, which is purely radiative for supertranslations.

However, the situation is different for charges associated with a Lorentz symmetry \(\xi = (X^a\hateq Y^a, \lambda = 0)\). In this case, the Maxwell fields contribute an additional term to the Wald-Zoupas charge given by
\be\label{eq:Q-EM-Y}
    \mc Q_{\rm{EM}}[Y; S] \defn \frac{1}{2\pi} \int_S\df \varepsilon_2~ \Re[\varphi_1] Y^a A_a.
\ee
We show in \cref{sec:examples} that this term vanishes for a Kerr-Newman black hole and thus does not affect the usual formula for its angular momentum. However, for general non-stationary Maxwell fields we expect that this term is non-vanishing. To illustrate this, we also consider a spinning charged sphere in Minkowski spacetime \cite{bpy}. The time-dependent dipole moment of such a charge distribution contributes non-trivially to \(\mc Q_{\rm EM}\) and thus to the angular momentum charge. A similar contribution to the angular momentum due to Maxwell fields is also present at spatial infinity in stationary-axisymmetric spacetimes \cite{SW1,SW2,KP-bundle}. Thus, the Maxwell contribution in \cref{eq:Q-EM-Y} would also be relevant to show that the Lorentz charges defined on future null infinity coincide with those defined at spatial infinity and at past null infinity, as conjectured in \cite{Stro-CK-match}.\\

\section{Discussion}
\label{sec:disc}

We analyzed the fluxes of Maxwell fields associated with the asymptotic symmetries at null infinity in any asymptotically flat spacetime. We first considered Maxwell theory in a non-dynamical background, defining three different currents which are naturally associated with vector fields on the background spacetime. When the vector field is a Killing vector field of the background spacetime, each of these currents is conserved, and they differ only by boundary terms. A similar situation occurs at null infinity when the vector field need not be a Killing vector field but an asymptotic symmetry element of the BMS algebra. In this case, each of the three currents can be used to construct fluxes associated with the asymptotic symmetry algebra through a given region of null infinity. While the Noether and canonical current fluxes are completely determined by the radiative degrees of freedom of the Maxwell fields, the flux associated with the asymptotic Lorentz symmetries defined by the stress-energy current also depends on the Coulombic part of the Maxwell field. Thus, if the stress-energy flux for a rotational symmetry is interpreted as the flux of angular momentum through null infinity, then it cannot be determined from the radiative degrees of freedom alone \cite{ab1,ab2}. Furthermore, none of these fluxes can be considered as the difference of charges evaluated on cross-sections of null infinity, as on a non-dynamical background spacetime, there is, in general, no notion of an energy or angular momentum of the Maxwell fields at a particular ``time'' defined by a cross-section of null infinity. Therefore, there is no obvious way to decide which of these currents defines the flux of energy or angular momentum.

To clarify this, we coupled electromagnetism to general relativity and considered the full Einstein-Maxwell theory at null infinity. Now, the theory is diffeomorphism-invariant, and there exist charges whose differences are given by fluxes. Specifically, the general prescription of Wald and Zoupas \cite{WZ} defines, for a given asymptotic symmetry, both the charge on a cross-section of \(\scri\) and the flux, which represents the change in this charge. If one assumes the charge expression for vacuum GR to be the definition of the charge in Einstein-Maxwell theory as well (see \cref{eq:Q-GR-defn}), then the additional term that Maxwell fields contribute to its flux is the stress-energy flux (\cref{eq:Q-GR-flux}). As in the case with a non-dynamical metric, this contribution depends on the Coulombic part of the Maxwell field for asymptotic Lorentz symmetries. However, the full Wald-Zoupas charge for Einstein-Maxwell theory contains an additional contribution to the charge due to the Maxwell fields (\cref{eq:Q-EM-defn}). This additional contribution vanishes for asymptotic supertranslations. It also vanishes for Lorentz symmetries in the Kerr-Newman spacetime. In general, however, for non-stationary Maxwell fields, this additional contribution is nonzero. The flux of the full Wald-Zoupas charge in Einstein-Maxwell theory with this additional contribution from Maxwell fields is determined by the radiative fields alone. The full Wald-Zoupas charge naturally absorbs the Coulombic information contained in the stress-energy flux, and so the contribution of the Maxwell fields to the Wald-Zoupas flux is determined by the Noether current flux and depends only on the radiative fields on \(\scri\). 

In addition, we showed, using the standard falloff conditions for the electromagnetic and gravitational fields near $i^0$ and $i^+$, that the Wald-Zoupas flux also defines a Hamiltonian generator associated with the asymptotic symmetries on all of null infinity.\\

A similar analysis can also be carried out for other matter fields. For GR minimally coupled to a massless Klein-Gordon field or a conformally-coupled scalar field, the essential points have already been discussed by Wald and Zoupas in Sec.~VI of \cite{WZ}. For such fields, the Wald-Zoupas charge is given by the same expression as in vacuum GR (\cref{eq:Q-GR-defn}) and the scalar fields contribute to the flux only through the stress-energy tensor. However, for Einstein-Yang-Mills theory, we expect that there is an additional contribution to the Wald-Zoupas charge similar to the case of Maxwell fields considered here.
For general theories, it should \emph{not} be expected that the matter contribution to the charge is the Noether charge or that the contribution to the flux is the Noether current. For instance, this expectation is already false in vacuum GR, where the Wald-Zoupas charge is, in general, \emph{not} given by the Noether charge (i.e., the Komar formula); see the discussion in \cite{GW,AW-linkage}.\\

As noted before, a similar additional contribution to the angular momentum due to Maxwell fields is also present at spatial infinity in stationary, axisymmetric spacetimes \cite{SW1,SW2,KP-bundle}. Thus, we expect that the Maxwell contribution in \cref{eq:Q-EM-Y} would also be relevant to show that the Lorentz charges defined on future null infinity coincide with those defined at spatial infinity and at past null infinity, as conjectured in \cite{Stro-CK-match}.
\\
Since the Wald-Zoupas flux is purely radiative and also the Hamiltonian generator on the radiative phase space of Einstein-Maxwell theory, it can also be quantized using the asymptotic quantization methods in \cite{asymp-quant}.\\

The Wald-Zoupas prescription can also be applied to finite null surfaces in vacuum GR \cite{CFP}. For Einstein-Maxwell theory at finite null surfaces, we expect that there is a similar contribution to the charges and fluxes associated with finite null boundary symmetries considered in \cite{CFP} that arises from the Maxwell fields. Such an analysis could also be useful in deriving conservation laws in Einstein-Maxwell theory through local regions bounded by a causal diamond similar to those in vacuum GR \cite{CP}.

\section*{Acknowledgements}

We thank Abhay Ashtekar for helpful discussions.
Research at Perimeter Institute is supported in part by the Government of Canada through the Department of Innovation, Science and Economic Development Canada and by the Province of Ontario through the Ministry of Economic Development, Job Creation and Trade. This work is also supported in part by NSF Grant No. PHY-1707800 to Cornell University.

\appendix

\section{Asymptotic symmetries of Einstein-Maxwell theory at null infinity}
\label{sec:asymp-symm}

In this appendix, we show how the asymptotic symmetries of Einstein-Maxwell theory can be derived from the asymptotic conditions on the gravitational and Maxwell fields at null infinity. We first focus on the asymptotic symmetries of the gravitational field, before we include the symmetry transformations of the Maxwell vector potential. Similar arguments for vacuum general relativity were also presented in \cite{FPS}.\\

Given a vector field \(\hat X^a = X^a\) generating an infinitesimal diffeomorphism  $\Lie_X \hat g_{ab}$ in the physical spacetime, what are the conditions on $X^a$ for it to be an asymptotic symmetry vector field?
The vector field $X^a$ needs to extend smoothly to $\scri$ to preserve the smooth differential structure there, and the infinitesimal diffeomorphisms generated by $X^a$ need to preserve the asymptotic flatness conditions on the unphysical metric perturbations. To make this concrete, consider any physical metric perturbation $\delta_X \hat{g}_{ab}= \Lie_{\hat X} \hat g_{ab}$ generated by a diffeomorphism. The corresponding unphysical metric perturbation is given by
\be\label{eq:unphys-diffeo}
    \delta_X g_{ab} = \Omega^2 \Lie_X \hat g_{ab} =  \Lie_X g_{ab} - 2 \Omega^{-1} n_c X^c g_{ab} \; .
\ee
Since \(\delta_X g_{ab}\) has to be smooth at \(\scri\), we can immediately conclude that \(n_a X^a \hateq 0\). In other words,  \(X^a\) is tangent to \(\scri\). 
Defining the function \(\alpha_{(X)} := \Omega^{-1} n_a X^a \), which extends smoothly to \(\scri\), we can write the above equation as
\be\label{eq:xi-cond}
	\delta_X g_{ab} = \Lie_X g_{ab} - 2 \alpha_{(X)} g_{ab} \; .
\ee

For the perturbation \(\delta_X g_{ab}\) to preserve the asymptotic flatness conditions in \cref{eq:asymp-flat-conds} and the Bondi condition in \cref{eq:Bondi-cond}, we require that
\be
    \delta_X g_{ab} \hateq 0 \qquad \text{and} \qquad n^a n^b \delta_X g_{ab} = O(\Omega^2) \; .
\ee
The first condition yields
\be\label{eq:conf-Kill}
    \Lie_X g_{ab} \hateq 2 \alpha_{(X)} g_{ab} \; .
\ee
Furthermore, contracting \cref{eq:xi-cond} with \(n^b\) gives
\be\label{eq:n-diffeo0}
    n^b\delta_X g_{ab} = n^b \nabla_b X_a - X^b \nabla_b n_a - \alpha_{(X)} n_a + \Omega \nabla_a \alpha_{(X)} \; ,
\ee
where we have used that the twist of $n_a$ vanishes, since $n_a$ is the gradient of the conformal factor \(\Omega\). Since the left-hand side must vanish at \(\scri\), we have
\be\label{eq:n-diffeo}
    n^b\delta_X g_{ab} \hateq 0 \implies
    \Lie_X n^a \hateq - \alpha_{(X)} n^a \; .
\ee
Contracting \cref{eq:n-diffeo0} once more with \(n^a\), we find that
\be\label{eq:nn-diffeo}
    n^a n^b \delta_X g_{ab} = O(\Omega^2) \implies \Lie_n \alpha_{(X)} \hateq 0 \; ,
\ee
where we used  \(n_{a}n^{a} = O(\Omega^{2})\) (see \cref{eq:nn-cond}, which followed directly from the Bondi condition in \cref{eq:Bondi-cond}).
Finally, taking the pullback of \cref{eq:conf-Kill} to $\scri$, we find
\be
    \Lie_X q_{ab} \hateq 2 \alpha_{(X)} q_{ab}\,.
\ee
Hence, the asymptotic symmetries on \(\scri\) are generated by vector fields \(X^a\) tangent to \(\scri\) satisfying 
\begin{subequations}\label[equation]{eq:bms-cond}\begin{align}
    \Lie_X n^a & \hateq - \alpha_{(X)} n^a \label{eq:xi-n-cond} \; , \\
    \Lie_X q_{ab} & \hateq 2 \alpha_{(X)} q_{ab} \label{eq:xi-q-cond} \; ,
\end{align}\end{subequations}
where the function \(\alpha_{(X)}\) is smooth and \(\Lie_n \alpha_{(X)} \hateq 0\) on \(\scri\). These conditions are the standard ones defining the BMS algebra \(\mf b\) \cite{Geroch-asymp,Ash-Str}. 
When working solely on $\scri$, the function \(\alpha_{(X)}\) can be interpreted as the infinitesimal conformal transformation of \(q_{ab}\) induced by \(X^a\vert_\scri\).
If \(X^a\) is given in a neighbourhood of \(\scri\), $\alpha_{(X)}$ can also be computed using
\be\label{eq:alpha-cond}
    \alpha_{(X)} \hateq \Omega^{-1} n_a X^a \hateq \tfrac{1}{4} \nabla_a X^a \; ,
\ee
where the second equality follows from \(g^{ab}\delta_X g_{ab} \hateq 0\).\\

To make these conditions more concrete, let \(u\) be an affine parameter along the null geodesics of \(n^a\) on \(\scri\) such that \(n^a \nabla_a u \hateq 1\). Then any BMS vector field can be written as
\begin{equation}\label{eq:BMS-decomp}
	 X^a \hateq \beta n^a + Y^a,
	 \qquad \text{with}\quad \beta \hateq f + \tfrac{1}{2}(u - u_0) \ms D_a Y^a,
\end{equation}
and
\be
\Lie_n f \hateq \Lie_n Y^a \hateq 0 \eqsp 2\ms D_{(a} Y_{b)} \hateq q_{ab} \ms D_c Y^c \; , \label{eq:lie-f}
\ee
where \(Y^a\) is tangent to the \(u = \text{constant}\) cross-sections of \(\scri\), \(\ms D_a\) is the covariant derivative on these cross-sections compatible with \(q_{ab}\), and \(u = u_0\) is some choice of an ``origin'' cross-section. The function \(\alpha_{(X)}\) in \cref{eq:bms-cond} is then given by \(\tfrac{1}{2} \ms D_a Y^a\) in this representation. Thus, any BMS vector field is characterized by a smooth function \(f\) and a smooth conformal Killing field \(Y^a\) on \(\bb S^2\). The function \(f\) represents the infinite-dimensional subalgebra of \emph{supertranslations} while the conformal Killing field \(Y^a\) represents a \emph{Lorentz} subalgebra of the full BMS Lie algebra.

Given a \emph{fixed} BMS vector field \(X^a\), its representation in terms of a supertranslation \(f\) and a Lorentz vector field \(Y^a\) depends on the choice of foliation given by \(u = \text{constant}\). Let \(u' = u + F\) with \(\Lie_n F \hateq 0\) be another choice of affine parameter along \(n^a\), and let \(f'\) and \({Y'}^a\) be representatives of \(X^a\) in the new choice of foliation given by \(u' = \text{constant}\). Then it is straightforward to verify that
\be\label{eq:bms-change}
    f' \hateq f + \Lie_Y F \eqsp {Y'}^a \hateq Y^a \; .
\ee
Therefore, the notion of a pure supertranslation (\(Y^a \hateq 0\)) is well-defined independently of the choice of foliation, but a ``pure Lorentz'' transformation (\(f=0\)) is not. This is ultimately related to the fact that the BMS algebra is a semidirect sum of the Lorentz algebra with the Lie ideal of supertranslations.\\

Now consider a similar analysis of the transformations of the Maxwell vector potential under a symmetry \(\xi = (X^a, \lambda)\), where \(X^a\) is a BMS vector field and \(\lambda = \hat \lambda\). The perturbation of the Maxwell vector potential generated by an infinitesimal transformation \(\xi\) is
\be
    \delta_\xi A_a = \Lie_X A_a + \nabla_a \lambda \; .
\ee
This transformation needs to preserve the asymptotic conditions of the Maxwell vector potential. Since \(A_a\) is smooth at \(\scri\), \(\lambda\) extends smoothly to \(\scri\) as well. To preserve the outgoing gauge condition imposed on the vector potential (\cref{eq:A_bondi}) requires that \(n^a \delta_\xi A_a \hateq 0\) which gives
\be\begin{aligned}
    & 0 \hateq n^a \Lie_X A_a + \Lie_n \lambda \\
    & \; \; \hateq \Lie_X (n^a A_a) + \alpha_{(X)} n^a A_a + \Lie_n \lambda \\
    & \implies \Lie_n \lambda \hateq 0 \; , \label{eq:lambda-dot}
\end{aligned}\ee
where the second line uses \cref{eq:xi-n-cond} and the last line follows from \(n^a A_a \hateq 0\). Thus, the asymptotic symmetries of Einstein-Maxwell theory at \(\scri\) are given by \(\xi = (X^a, \lambda)\), where \(X^a\) is a BMS vector field and \(\lambda\) is any smooth function on \(\bb S^2\), the space of null generators of \(\scri\).

Similar to the case of a BMS vector field, the representation of a \emph{fixed} \(\xi\) in terms of a BMS vector field \(X^a\) and a Maxwell gauge transformation \(\lambda\) depends on the choice of gauge for the background vector potential \(A_a\). Let \(A'_a = A_a + \nabla_a \Lambda\) be another vector potential related to \(A_a\) by a gauge transformation \(\Lambda\) with \(\Lie_n \Lambda \hateq 0\). For a fixed symmetry \(\xi = (X^a, \lambda)\) let the new representatives under the gauge transformation by \(\Lambda\) be \(\xi = ({X'}^a, \lambda')\). Since the symmetry \(\xi\) is fixed, its action on the vector potentials must be independent of the choice of gauge, that is, \(\delta_\xi A'_a = \delta_\xi A_a\). Evaluating this, we have
\be\begin{aligned}
    \Lie_{X'} A_a + \nabla_a \lambda' + \nabla_a \Lie_{X'} \Lambda &= \Lie_X A_a + \nabla_a \lambda \; .
\end{aligned}\ee
This implies that under a change of Maxwell gauge by \(\Lambda\) the representation of a fixed symmetry \(\xi = (X^a, \lambda) = ({X'}^a, \lambda')\) changes as
\be\label{eq:symm-gauge}
    {X'}^a = X^a \eqsp \lambda' = \lambda - \Lie_X \Lambda \; .
\ee
Consequently, the notion of a pure Maxwell gauge transformation \(\xi = (X^a = 0, \lambda)\) is well-defined independently of the choice of gauge \(\Lambda\), but a ``pure BMS transformation'' \(\xi = (X^a, \lambda = 0)\) is not. This is analogous to the structure of the BMS algebra noted above. Note also that
\be\label{eq:inv}
    \lambda' + {X'}^a A'_a = \lambda + X^a A_a
\ee
is invariant under changes of Maxwell gauge.\footnote{In the principal bundle picture, where \(\xi = (X^a, \lambda)\) is a vector field on the bundle, the Lie algebra of such vector fields also has the structure of a semidirect sum of diffeomorphisms with the Lie ideal of Maxwell gauge transformations \cite{KP-bundle}. The invariant in \cref{eq:inv} is then the vertical part of \(\xi\) on the bundle.}

\section{Stationary solutions in Einstein-Maxwell theory at null infinity}
\label{sec:stationary}

In this appendix, we show that for any stationary solution \((\hat g_{ab}, \hat A_a)\) of Einstein-Maxwell theory, which is asymptotically flat, the radiative field \(\mc E_a\) and the News tensor \(N_{ab}\) vanish at \(\scri\).
To do so, we will first show that any nonzero timelike Killing vector field \(\hat t^a\) in the unphysical spacetime is necessarily a nonzero supertranslation on \(\scri\).\footnote{It can further be shown that the timelike Killing field is a BMS translation (see Lemma~1.4 of \cite{AX} and also p.~54 of \cite{Geroch-asymp}), but we will not need this stronger result.} Then, we show that this implies that \(\mc E_a = 0\) on \(\scri\) for any solution of the Maxwell equation which is stationary, i.e., \(\Lie_{\hat t} \hat F_{ab} = 0\). Finally, using the proof by Geroch \cite{Geroch-asymp}, this also implies that \(N_{ab} = 0\).\\

On \(\scri\), a supertranslation vector field takes the form \(X^a \hateq f n^a\) with \(\Lie_n f \hateq 0\). For our purposes we will also need the ``subleading'' form of this vector field away from \(\scri\); see, for instance, Eq.~21 of \cite{GW} and Eq.~93 of \cite{WZ}. For completeness, we collect the proof in the following lemma.

\begin{lemma}\label{lem:st-form}
    Any vector field \(X^a\) in \(M\) such that \(X^a\vert_\scri\) is a BMS supertranslation is of the form
\be\label{eq:st-inside}
    X^a = f n^a - \Omega \nabla^a f + O(\Omega^2)
\ee
for some \(f\) smooth in \(M\) and \(\Lie_n f \hateq 0\).
\begin{proof}
    Since \(X^a\vert_\scri\) is a BMS supertranslation, we have \(X^a \hateq f n^a\) for some \(f\) on \(\scri\) satisfying \(\Lie_n f \hateq 0\). Now extend the function \(f\) \emph{arbitrarily} but smoothly into \(M\), and thus $X^a$ takes the form
    \be
        X^a = f n^a + \Omega Z^a
    \ee
    for some smooth \(Z^a\). Then, using \cref{eq:nn-cond,eq:alpha-cond}, \(\alpha_{(X)} \hateq n_a Z^a\). Using the Bondi condition (\cref{eq:Bondi-cond}), \cref{eq:conf-Kill} for such a vector field becomes
    \be
        \nabla_{(a} f n_{b)} + n_{(a} Z_{b)} \hateq n_c Z^c g_{ab} \; .
    \ee
    Taking the trace on both sides gives \(n_a Z^a \hateq 0\), and consequently \(Z_a = - \nabla_a f \). 
\end{proof}
\end{lemma}
Note that we extended the function \(f\) away from \(\scri\) in an arbitrary manner. It is easy to check from \cref{eq:st-inside} that the freedom in this extension affects only the \(O(\Omega^2)\) part of the vector field. One can choose to fix the \(O(\Omega^2)\) part by choosing some convenient choice of conformal factor and coordinates (such as Bondi coordinates) away from \(\scri\), but we will not need to do so.\\

Now we turn to timelike Killing fields of the physical spacetime \((\hat M, \hat g_{ab})\) and show that they correspond to nontrivial supertranslations on null infinity.

\begin{lemma}\label{lem:t-is-st}
    Let \(\hat t^a\) be a nonzero timelike Killing vector field in the physical spacetime \((\hat M, \hat g_{ab})\). Then \(t^a = \hat t^a\) is a nonzero supertranslation on \(\scri\).
\begin{proof}
    Since \(\Lie_{\hat t} \hat g_{ab} = 0\), from \cref{eq:unphys-diffeo} it follows that \(t^a = \hat t^a\) is a BMS vector field on \(\scri\). Since \(\hat t^a\) is timelike in the physical spacetime, we have \(\hat g_{ab} \hat t^a \hat t^b < 0\). In the unphysical spacetime away from null infinity (i.e., on \(M-\scri\)), this gives \(\Omega^{-2} g_{ab} t^a t^b < 0\). Now \(\Omega > 0\) on \(M - \scri\), \(\Omega \hateq 0\), and \(g_{ab}\) and \(t^a\) extend smoothly to \(\scri\), and thus
    \be
    g_{ab} t^a t^b \leq 0
    \ee
    in \(M\), with the equality possibly holding on \(\scri\). Writing \(t^a \hateq \beta n^a + Y^a\) (from \cref{eq:BMS-decomp}), we get that \(q_{ab} Y^a Y^b \leq 0\) on \(\scri\).
Since \(q_{ab}\) is a Riemannian metric on the cross-sections of \(\scri\) and \(Y^a\) is tangent to these cross-sections, this means \(Y^a \hateq 0\). Thus the ``Lorentz part'' of \(t^a\) vanishes and \(t^a\) is a BMS supertranslation.

    Next, we show that this supertranslation is necessarily nonzero on \(\scri\) (see also \cite{AX}). We will proceed by assuming that \(t^a \hateq 0\) and show that this implies that \(\hat t^a\) vanishes everywhere, contradicting the assumption that it is a nonzero Killing vector field. Since \(t^a\) is a supertranslation on \(\scri\), if \(t^a \hateq 0\), then from \cref{lem:st-form} we have that
    \be\label{eq:t-if-trivial}
        t^a = \Omega^2 W^a,
    \ee
    for some smooth \(W^a\). Since \(\hat t^a\) is a Killing vector field in the physical spacetime \((\hat M, \hat g_{ab})\), \(t^a\) is a conformal Killing field in the unphysical spacetime \((M, g_{ab})\) with
\be
    \Lie_t g_{ab} = 2 \alpha_{(t)} g_{ab} \eqsp \alpha_{(t)} = \Omega^{-1}n_a t^a \; .
\ee
Any conformal Killing field is completely determined by its \emph{conformal Killing data} specified at any point \(p \in M\) \cite{AMA-isometries}:
    \be
    \lb( t^a, \nabla_{[a} t_{b]}, \alpha_{(t)}, \nabla_a \alpha_{(t)} \rb) \big\vert_p \; .
    \ee
    Furthermore, if the conformal Killing data vanish at any point \(p\), then the conformal Killing field \(t^a\) vanishes \emph{everywhere}. We now show that the conformal Killing data of \cref{eq:t-if-trivial} vanish on \(\scri\). It is easy to see by a direct computation that \(t^a\), \(\nabla_{[a} t_{b]}\), and \(\alpha_{(t)}\) vanish on \(\scri\). Computing the remaining last piece of the conformal Killing data, we have
    \be
    \nabla_a \alpha_{(t)} \hateq n_a (n_b W^b) \; .
    \ee
    To show that this vanishes at \(\scri\), we evaluate \(\Lie_t g_{ab} = 2 \alpha_{(t)} g_{ab}\) with \cref{eq:t-if-trivial} to get
    \be
    4 \Omega n_{(a} W_{b)} + 2 \Omega^2 \nabla_{(a} W_{b)} = 2 \Omega n_c W^c g_{ab} \; .
    \ee
    Note that this holds in a neighbourhood of \(\scri\) and not just on \(\scri\), as a consequence of \(\hat t^a\) being Killing in the physical spacetime. Multiplying the above equation by \(\Omega^{-1}\), taking the trace, and then taking the limit to \(\scri\), we get \(n_a W^a \hateq 0\), and so \(\nabla_a \alpha_{(t)} \hateq 0\). Hence, all the conformal Killing data for the conformal Killing field of the form \cref{eq:t-if-trivial} vanish on \(\scri\), and thus \(t^a = 0\) everywhere in \(M\). This implies that \(\hat t^a = 0\) in \(\hat M\), which contradicts the assumption that \(\hat t^a\) is a nonzero Killing field in the physical spacetime. Thus, any nonzero timelike Killing vector field in the physical spacetime is necessarily a nonzero supertranslation on \(\scri\).
\end{proof}
\end{lemma}

Finally, we show that, for a stationary solution of Einstein-Maxwell theory, the radiative fields \(N_{ab}\) and \(\mc E_a\) vanish on null infinity.\footnote{Note that for this result to hold it is essential that the space of generators of \(\scri\) is topologically \(\bb S^2\).}

\begin{thm}
Let \((\hat g_{ab}, \hat A_a)\) be a stationary solution of Einstein-Maxwell theory, that is, there exists a timelike vector field \(\hat t^a\) in the physical spacetime \(\hat M\) such that
\be
    \Lie_{\hat t} \hat g_{ab} = 0 \qquad \text{and} \qquad \Lie_{\hat t} \hat F_{ab} = 0 \; .
\ee
Then, the radiative fields vanish on $\scri$: \(N_{ab} \hateq 0\) and \(\mc E_a \hateq 0\).
\begin{proof}
Consider first the stationary electromagnetic field \(\hat F_{ab}\), for which in the unphysical spacetime we have \(\Lie_t F_{ab} = 0\), where as before \(t^a = \hat t^a\). From \cref{lem:st-form,lem:t-is-st}, we have that
\be\label{eq:t-form}
    t^a = f n^a - \Omega \nabla^a f +O(\Omega^2)
\ee
for some \(f \neq 0\) and \(\Lie_n f \hateq 0\). Evaluating the pullback of \(\Lie_t F_{ab}n^b = 0\) to \(\scri\) and using \(\Lie_t n^a \hateq 0\) and \(\Lie_n f \hateq 0\) (as $t^a$ is a supertranslation) gives
\be\label{eq:Lie-n-E}
    \Lie_n (f \mc E_a) \hateq 0.
\ee
Similarly, evaluating the pullback of \(\Lie_t F_{ab} = 0\) to \(\scri\), we have
\be\label{eq:E-curl}
    \ms D_{[a} (f \mc E_{b]}) \hateq 0.
\ee
Note that only the derivative along the cross-sections \(\ms D_a\) occurs in this equation due to \cref{eq:Lie-n-E} and the Bondi condition (\cref{eq:Bondi-cond}). Next, evaluating \(l^a n^b \Lie_t F_{ab} \hateq 0\), we have
\be\label{eq:E-div}\begin{aligned}
    0 & \hateq l^a n^b \Lie_t F_{ab} \hateq \Lie_t (F_{ab}l^a n^b) - F_{ab} \Lie_t l^a n^b \\
    & \hateq f \Lie_n (F_{ab}l^a n^b) + F_{ab} (n^a \Lie_l f  + \nabla^a f) n^b \\
    & \hateq f q^{ab} \ms D_a \mc E_b + q^{ab} \mc E_a \ms D_b f \\
    & \hateq q^{ab} \ms D_a(f \mc E_b),
\end{aligned}\ee
where the first line uses \(\Lie_t n^a \hateq 0\) for a supertranslation, the second line is a straightforward computation using \cref{eq:t-form}, and the third line uses the Maxwell equation \cref{eq:n_maxwell}. From \cref{eq:Lie-n-E,eq:E-curl,eq:E-div}, it follows that \(f \mc E_a\) is a covector field on the space of generators of \(\scri\) with vanishing curl and divergence. Since the space of generators of \(\scri\) is topologically \(\bb S^2\) and $f \neq 0$, this implies that \(\mc E_a = 0\) for any stationary solution.

Now, from \cref{eq:T-Maxwell}, we have that \(T_{ab}n^an^b \hateq \tfrac{1}{4\pi} \mc E_a \mc E^a\), and thus for any stationary solution \(T_{ab}n^an^b \hateq 0\). With this condition and the Einstein equation, it can be shown that $N_{ab} \hateq 0$ for any stationary spacetime (see pp.~53--54 of \cite{Geroch-asymp}). Thus, for any stationary solution of the Einstein-Maxwell equations, we have \(N_{ab}\hateq 0 \hateq \mc E_a \), as we wished to show.
\end{proof}
\end{thm}

\section{Computation of \(\mc Q_{\rm EM}\) in some examples}
\label{sec:examples}

In this appendix, we give two examples of the Maxwell contribution to the Wald-Zoupas charge $\mc Q_{\rm EM} [Y; S]$ of an asymptotic Lorentz symmetry $Y^a$. This contribution vanishes for the first example of Kerr-Newman spacetimes, while for the second example of a spinning charged sphere with variable angular velocity it is nonzero.

\subsection{Kerr-Newman spacetime}

The line element of the (physical) Kerr-Newman metric in Boyer-Lindquist coordinates $(t,r,\theta,\phi)$ is given by (see Appendix~D.1 of \cite{Frolov-Novikov})
\begin{align}
 d s^2 &= - \left( 1 - \frac{2 M r - Q^2}{\Sigma}\right) dt^2
 - \frac{2 a \sin^2 \theta (2 M r - Q^2)}{\Sigma} dt d\phi
 + \frac{\Sigma}{\Delta} d r^2 + \Sigma \; d \theta^2  \notag \\
 & \qquad + \left( (r^2 + a^2)^2 - a^2 \sin^2 \theta \; \Delta \right)\frac{\sin^2 \theta}{\Sigma}d \phi^2,
\end{align}
with
\begin{equation}
	\Sigma := r^2 + a^2 \cos^2 \theta \qquad \text{and} \qquad \Delta := r^2 - 2 M r + a^2 + Q^2.
\end{equation}
Since we wish consider the limit to \(\scri\), it is more convenient to introduce the \emph{outgoing} null coordinates \(x^\mu = (u,r,\theta,\phi)\), with \(u\) defined by
\be
    d u = d t - \frac{r^2 + a^2}{\Delta}  d r \;.
\ee
The (physical) Kinnersley tetrad --- normalized such that $\hat{l}^\mu \hat{n}_\mu = -1$ and $\hat{m}^\mu \hat{\bar{m}}_\mu=1$ --- in these coordinates is
\begin{subequations}
	\begin{align}
	\hat{l}^\mu \partial_\mu & = \partial_r + \frac{a}{\Delta} \partial_\phi, \\
	\hat{n}^\mu \partial_\mu & = \frac{r^2+a^2}{\Sigma}\partial_u
	- \frac{\Delta}{2 \Sigma}  \partial_r  + \frac{ a}{2 \Sigma} \partial_\phi, \\
	\hat{m}^\mu \partial_\mu & = \frac{i a \sin \theta}{\sqrt{2} (r+ i a \cos \theta)} \partial_r + \frac{1}{\sqrt{2} (r + i a \cos \theta)} \Big( \partial_\theta  +  \frac{i }{\sin \theta} \partial_\phi \Big)  \; .
	\end{align}
\end{subequations}

The Maxwell vector potential in these null coordinates is
\begin{equation}\label{eq:A-KN}
\hat{A}_\mu dx^\mu = - \frac{r Q}{\Sigma} \left( d u +
\frac{r^2 + a^2}{\Delta}  d r
- a \sin^2 \theta d \phi  \right)  \; ,
\end{equation}
which satisfies the Lorenz gauge condition \(\hat g^{\mu\nu} \hat\nabla_\mu \hat A_\nu = 0\).\\

To take the limit to \(\scri\), we use the conformal factor $\Omega = r^{-1}$ and use \(\Omega\) as the new coordinate instead of \(r\). It can be verified that the unphysical metric \(g_{\mu\nu} = \Omega^2\hat g_{\mu\nu}\) is smooth in the limit to \(\scri\) (that is, as $\Omega \to 0$ with fixed $u,\theta,\phi$). The unphysical tetrad \((l^\mu, n^\mu, m^\mu, \bar m^\mu)\) defined by
\begin{subequations}
	\begin{align}
	l^\mu \partial_\mu & \defn \Omega^{-2}  \hat{l}^\mu \partial_\mu = \partial_\Omega + O(\Omega), \\
	n^\mu \partial_\mu & \defn \hat{n}^\mu \partial_\mu  =  \partial_u + O(\Omega), \\
	m^\mu \partial_\mu & \defn \Omega^{-1} \hat{m}^\mu \partial_\mu = \tfrac{1}{\sqrt{2}} \lb( \partial_\theta + \tfrac{i }{\sin \theta} \partial_\phi \rb) + O(\Omega)\; , 
\end{align}
\label{eq:m-unphys}
\end{subequations}
is also smooth at \(\scri\). The unphysical \(n^\mu\) defined above coincides with the normal \(n^a = g^{ab}\nabla_b \Omega\) at \(\scri\) to leading order, but not at $O(\Omega)$, as this \(n^\mu\) does not satisfy the Bondi condition.

The vector potential \(A_\mu = \hat A_\mu\) in \cref{eq:A-KN} is not smooth at \(\scri\), since \(l^\mu A_\mu\) diverges as \(\Omega \to 0\). However, instead, consider the vector potential \(A'_\mu\) related to \cref{eq:A-KN} by a gauge transformation:
\be\label{eq:newA-KN}
    A_\mu' =  A_\mu - \nabla_\mu (Q \ln \Omega).
\ee
This new vector potential \(A'_\mu\) is no longer in Lorenz gauge (in the physical spacetime) but is smooth at \(\scri\), and it also satisfies the outgoing radiation gauge condition $n^\mu A'_\mu \hateq 0$. Henceforth, we use this smooth vector potential on \(\scri\) and drop the ``prime'' from the notation.

On \(\scri\), the Lorentz vector fields \(Y^a\) are spanned by the tetrads \(m^\mu\) and \(\bar m^\mu\). A direct computation using \cref{eq:A-KN,eq:m-unphys,eq:newA-KN} gives \(m^\mu A_\mu \hateq 0 \) and consequently \(Y^a A_a \hateq 0\) for all Lorentz vector fields. Thus, in the Kerr-Newman spacetime, the Maxwell contribution to the Lorentz charges vanishes; i.e., $\mc Q_{\rm EM} [Y; S] =0$. In particular, the angular momentum of the Kerr-Newman black hole computed using the Wald-Zoupas charge (with \(Y^a \equiv \partial_\phi\)) gets no additional contribution from \(\mc Q_{\rm EM}\) and is thus given by the standard result (see, for example, \cite{Winicour1965}).\footnote{To calculate the Wald-Zoupas charge using \cref{eq:Q-GR-defn}, one needs to be careful to use a tetrad where the \(n^a\) satisfies the Bondi condition \cref{eq:Bondi-cond,eq:nn-cond} and the corresponding \(l^a\), and not the tetrad in \cref{eq:m-unphys}.
}

\subsection{Spinning charged sphere in Minkowski spacetime}
\label{sec:charged-sphere}

The above computation of the Lorentz charges in Kerr-Newman spacetimes does not mean that the electromagnetic contribution to the Wald-Zoupas charge for angular momentum will always vanish. An explicit example for which $\mc Q_{\rm EM}$ is nonzero is considered in \cite{bpy}: a thin spherical shell in Minkowski spacetime, with radius $R$ and charge $Q$, spinning on a central axis with a time-dependent angular velocity \(\omega(t)\). The time-dependent dipole moment of the spherical shell is given by $d(t) = \tfrac{1}{3} Q R^2 \omega(t)$. Furthermore, \cite{bpy} also assumes that the characteristic timescale of variation of the magnetic dipole moment is much greater that the light-travel time \(\tau = R\) across (half) the sphere, that is,
\be\label{eq:slow-rot}
    \frac{\partial}{\partial t} d(t) \ll \frac{d(t)}{\tau}.
\ee
This is clearly not a solution to the source-free Maxwell equations due to the presence of a source current. However, given that the source current is compact, our analysis in the main body of the paper still applies. 
We do not attempt to solve the full Einstein-Maxwell equations for this system. Thus, the Maxwell field in this section should be thought of as a perturbation generated by the charged sphere on the background Minkowski spacetime.\\

The relevant null tetrads at \(\scri\) in Minkowski spacetime can be constructed in the same manner as in the Kerr-Newman spacetime by taking \(M=a=Q = 0\). To get a smooth vector potential at \(\scri\), one again needs to perform a gauge transformation as in \cref{eq:newA-KN} which takes us out of the Lorenz gauge used in \cite{bpy}. Then, from the explicit computations in \cite{bpy}, it can be shown that
\be\label{eq:charged-sphere-fields}
    \Re[\varphi_1] \hateq \frac{1}{2}Q \eqsp m^a A_a \hateq \frac{i}{\sqrt{2}}  \Gamma^{(0)}(u) \sin \theta \, ,
\ee
where $u= t-r$ is the retarded time coordinate and we have taken the rotation axis for the sphere to be along the $z$-axis. With the assumption \cref{eq:slow-rot}, the function \(\Gamma^{(0)}(u)\) is given by
\begin{equation}\label{eq:Gamma-thing}
\Gamma^{(0)}(u) := \frac{\partial}{\partial u} d(u) + \frac{1}{10} \tau^2 \frac{\partial^3}{\partial u^3} d(u) + \frac{1}{280} \tau^4 \frac{\partial^5}{\partial u^5} d(u) + \cdots,
\end{equation}
where \(\cdots\) denotes higher-order terms.

Now, a rotational Killing vector field along the $z$-axis is given by $R_{(z)}^a = - \tfrac{i}{2} \sin \theta \; (m^a - \bar m^a)$. Thus, using \cref{eq:charged-sphere-fields,eq:Gamma-thing}, we can compute the Maxwell contribution to the charge of \(R_{(z)}^a\) (\cref{eq:Q-EM-defn}) --- the angular momentum in the \(z\) direction --- on a \(u=\text{constant}\) cross-section \(S_u\) to be
\begin{align}
    \mc Q_{\rm EM}[ R_{(z)};S_u]  = \frac{\sqrt{2}}{3} Q ~ \Gamma^{(0)}(u) \; .
\end{align}

Thus, we expect that generic non-stationary Maxwell fields will contribute a non-vanishing \(\mc Q_{\rm EM}\) to the Wald-Zoupas charge for asymptotic Lorentz symmetries.



\bibliographystyle{JHEP}
\bibliography{EM-currents}
\end{document}